\documentclass[pas,manuscript,showpacs,showkeys,superscriptaddress,nofootinbib]{revtex4-1}
\pdfoutput=1
\usepackage{graphicx}
\usepackage{epsfig}  
\usepackage{epsf}    
\usepackage[tbtags]{amsmath}
\usepackage{amsfonts}
\usepackage{filecontents}
\usepackage{float}
\usepackage{makecell}
\usepackage{subfig}
\usepackage{dcolumn}
\usepackage{bm}
\usepackage{dcolumn}
\usepackage{textcomp}
\usepackage{multirow}
\usepackage{slashed}
\usepackage{float}
\restylefloat{figure}
\usepackage[]{hyperref}
  \hypersetup{
  bookmarks=true,         
  unicode=true,         
  pdftoolbar=true,     
  pdfmenubar=true,     
  pdffitwindow=true,    
  pdfstartview={FitH},    
  pdfsubject={Sterilenus@DUNE},  
  pdfnewwindow=true,     
  pdfcreator={RevTeX},
  colorlinks=true,     
  linkcolor=red,   
  citecolor=blue,    
  filecolor=black,  
  urlcolor=blue,           
  }
\usepackage{hypcap}
\def\nue{{\nu_e}}
\def\anue{{\bar{\nu}_e}}
\def\numu{{\nu_{\mu}}}
\def\anumu{{\bar{\nu}_{\mu}}}

\newcommand{\eg}{{\it e.g.}}
\newcommand{\ie}{{\it i.e.}}

\newcommand{\beq}{\begin{equation}}
\newcommand{\eeq}{\end{equation}}
\newcommand{\beqa}{\begin{eqnarray}}
\newcommand{\eeqa}{\end{eqnarray}}

\usepackage{cancel}
\usepackage{soul}

\begin{document}
\title{Understanding the MiniBooNE  and the muon and electron $g-2$ anomalies  with a light $Z'$ and a second Higgs doublet \vspace{-0.15cm}}
\author{Waleed Abdallah}
\email[Email Address: ]{waleedabdallah@hri.res.in}
\affiliation{Harish-Chandra Research Institute, HBNI, Chhatnag Road, Jhunsi, Allahabad 211019, India\looseness=-1}
\affiliation{Department of Mathematics, Faculty of Science, Cairo University, Giza 12613, Egypt\looseness=-1}
\author{Raj Gandhi}
\email[Email Address: ]{raj@hri.res.in}
\affiliation{Harish-Chandra Research Institute, HBNI, Chhatnag Road, Jhunsi, Allahabad 211019, India\looseness=-1}
\author{Samiran Roy}
\email[Email Address: ]{samiran@prl.res.in}
\affiliation{Harish-Chandra Research Institute, HBNI, Chhatnag Road, Jhunsi, Allahabad 211019, India\looseness=-1}
\affiliation{Physical Research Laboratory, Ahmedabad - 380009, Gujarat, India}
\begin{abstract}
Two of the most widely studied extensions of the Standard Model (SM) are $a)$ the addition of a new $U(1)$ symmetry to its existing gauge groups, and $b)$ the expansion of its scalar sector to incorporate a second Higgs doublet. We show that when combined, they allow us to understand  the electron-like event excess seen in the MiniBooNE (MB) experiment as well as account for  the observed  anomalous values of the muon magnetic moment. A light $Z'$ associated with an additional $U(1)$ coupled to baryons and to the dark sector, with flavor non-universal couplings to leptons,  in conjunction with a second Higgs doublet is capable of explaining the MB excess.  The $Z'$ obtains its mass from a dark singlet scalar, which mixes with the two Higgs doublets. Choosing  benchmark parameter values, we show that $U(1)_{B-3L_\tau}$, which is anomaly-free,  and $U(1)_B$, both provide (phenomenologically) equally good solutions to the excess.  We  also point out the other (anomaly-free) $U(1)$ choices that may be  possible upon fuller exploration of the parameter space.  We obtain very good matches to the energy and angular distributions for neutrinos and anti-neutrinos in  MB. The extended Higgs sector has two light CP-even scalars, $h'$ and $H$, and their masses and couplings are such that in principle, both contribute to  help explain  the MB excess as well as the  present observed values of the muon and electron $g-2$. We  discuss the constraints on our model as well as future tests.  Our work underlines the role that light scalars may play in understanding present-day low-energy anomalies. It also  points to the  possible existence of  portals to the dark sector, \textit{i.e.}, a light gauge boson field $(Z')$ and a dark neutrino which mixes with the active neutrinos, as well as a dark sector  light scalar which mixes with the extended Higgs sector.
\end{abstract}
\keywords{\vspace{-0.45cm}MiniBooNE excess, Muon and electron $g-2$, $U(1)_{B}$, $U(1)_{B-3L_{\tau}}$}
\maketitle
\section{Introduction}
\label{sec1}
The Standard Model (SM) of particle physics with its underlying framework of local gauge symmetries~\cite{Tanabashi:2018oca}\footnote{For detailed pedagogical treatments see, for instance,~\cite{Quigg:2013ufa,Pal}.} is  a highly successful present-day theory. It  explains, with impressive accuracy,  an unprecedented range of experimental measurements  over many decades in energy. In spite of its stellar success, however, the list of reasons as to why physics beyond the Standard Model (BSM) should  exist is both long and compelling. Dark matter (DM)~\cite{Arun:2017uaw,Kahlhoefer:2017dnp,Gaskins:2016cha,  Bertone:2004pz,Feng:2010gw},  the  existence of which is extensively supported by a range of astronomical observations, is one of the strongest motivations for looking  for new physics, because it is clear that none of the SM particles can contribute significantly to its share of the energy density of our universe.  It is fair to say that despite assiduous efforts, practically no light has been shed  so far on its particle properties. 

\vspace{0.2cm}
The observed matter and anti-matter asymmetry in our universe~\cite{Pascoli:2020swq, Tanabashi:2018oca, Canetti:2012zc} and the existence of small but non-zero neutrino mass differences~\cite{PhysRevLett.87.071301,PhysRevLett.81.1562,PhysRevD.88.032002,PhysRevLett.108.191802},  with masses widely different in magnitude from those  of the charged leptons and quarks, as well as the existence of three families of quarks and leptons with a large mass hierarchy provide further grounds for the search for BSM physics. 

\vspace{0.2cm}
A puzzling, and to a degree, unanticipated development in the effort to discover new physics is the lack of any 
definitive signals  pointing to  its presence at the Large Hadron Collider (LHC). Most notable among these is the absence (so far) of  supersymmetry~\cite{Martin:1997ns}, which, arguably,  has been  the most popular model for BSM physics over the last three decades. This has led to renewed interest in the quest for BSM signals in other experiments, in settings as diverse as  B-factories, rare decay searches, muon storage rings, matter-wave interferometers, pair-spectrometers for nuclear transitions and neutrino and DM detectors.

\vspace{0.2cm}
These efforts have not been disappointing. At the present time, there are several empirical results which appear to be anomalous at levels of statistical significance which invite, and in some cases, demand attention. Among them  are  observed discrepancies in $a)$ the values of the anomalous magnetic moment of the muon~\cite{Miller:2007kk,Aoyama:2020ynm}  and the electron~\cite{Parker:2018vye}, $b)$ excesses in electron events in tension with muon neutrino disappearance data at short-baseline neutrino detectors~\cite{Maltoni:Talk}, $c)$ a significant excess in the signal versus background expectation in the KOTO experiment~\cite{Ahn:2018mvc} which searches for the decay of a neutral kaon to a neutral pion and a neutrino pair, $d)$ discrepancies with SM predictions in observables related to B-decays~\cite{London:2019nlu}, and finally, $e)$ anomalies in the decay of excited states of Beryllium~\cite{DelleRose:2018pgm}.

\vspace{0.3cm}
The possibility of connections between two or more of the sectors motivating the search for BSM physics  has generated significant interest of late, and  this work is also based on such a connection. For instance, the connection between  neutrinos and the dark sector\footnote{In what follows, the dark sector is assumed to comprise of particles which do not couple to SM fermions or gauge bosons, or do so extremely weakly and indirectly, \textit{e.g.} via kinetic or mass mixings.} pursued here has recently been discussed in~\cite{1103.3261, 1202.6073, 1203.0545, 1604.06099, 1812.05102, 1903.00006, 1903.07589, 1907.08311}. 

\vspace{0.3cm}
For our purpose here, we note that if DM interacts with particles of the SM, its scattering  must resemble neutral current interactions of neutrinos. This similarity is the reason why coherent elastic neutrino-nucleon  scattering (CE${\nu}$NS)~\cite{PhysRevD.9.1389,Kopeliovich:1974mv} is a major background for next generation DM experiments looking to directly detect weakly interacting massive particles (WIMPs)~\cite{1307.5458}. This correspondence also underlies proposals and  sensitivity studies for the direct detection of   DM at fixed-target neutrino experiments (see, \textit{e.g.}~\cite{1107.4580,1205.3499,1211.2258,1405.7049,1505.07805,1609.01770,1702.02688,1807.06137, 1807.06501}) or even at much higher energies~\cite{1407.3280,1503.02669,1612.02834,Arguelles:2019boy}. It follows, therefore, that persistent anomalous excesses in neutrino experiments should be scrutinized keeping in mind that they may be receiving contributions from dark sector particles scattering off SM particles via a mediating portal particle,  which could be $i)$ a vector, $ii)$ a scalar or $iii)$ a dark neutrino which mixes with the SM neutrinos.

\vspace{0.3cm}
In this work, we propose a solution to  the electron-like event excess seen in the MiniBooNE (MB) experiment based on a new $U(1)$ symmetry associated with  baryon number, mediated by a  light new neutral gauge boson $Z'$, which couples either selectively or not at all to leptons. It also couples directly  to particles in the dark sector and indirectly to neutrinos, via mixing. We do not propose a unique choice for the new gauge group insofar as its coupling to SM particles is concerned, but via benchmark parameters, show that both  $U(1)_{B-3L_\tau}$ and $U(1)_B$ provide equally good solutions to the excess. We also  indicate other (anomaly-free) choices that may be allowed once the parameter space is fully explored. The interaction (described in more detail below) which leads to the observed MB excess involves a dark neutrino,  $\nu_d$, mixed with the SM neutrinos, a SM Higgs sector expanded to include a second doublet,  and a singlet (under the SM) scalar which couples to the  SM fermions only via its mass mixing with the two Higgs doublet (2HD) sector\footnote{A more economical possibility, where only a singlet scalar with mass mixing to the SM Higgs is added, is precluded by very tight constraints on its fermionic couplings from a variety of experiments, see~\cite{Winkler:2018qyg}.}. While providing a very good fit to the MB data, this also accounts for the present observed value of  the anomalous muon  magnetic moment, without further embellishment or fine-tuning.

Section~\ref{sec2} discusses the observed excess in  MB and the measured discrepant value of the  anomalous muon and electron magnetic moments. Section~\ref{sec3} discusses our model, its motivations and Lagrangian, and presents the calculation of the process that leads to our explanation of the MB excess. Section~\ref{sec4-A} presents our results for  MB and compares the neutrino and anti-neutrino energy and angular distributions obtained with the data. Sections~\ref{sec4-B} and~\ref{sec4-C} focus on the implications of our model for the anomalous magnetic moment of the muon and electron, respectively. Section~\ref{sec5} focuses on  constraints on our work and discusses some possible future tests. Section~\ref{sec6} qualitatively summarizes our results and conclusions.
\vspace{-0.5cm}
\section{The MiniBooNE/LSND, the muon and electron $g-2$ anomalies}
\label{sec2}
\subsection{Event excesses in MiniBooNE and LSND}
It is well-known that two neutrino experiments, MiniBooNE (MB)~\cite{AguilarArevalo:2007it,AguilarArevalo:2008rc,AguilarArevalo:2010wv,Aguilar-Arevalo:2013pmq, Aguilar-Arevalo:2018gpe,Aguilar-Arevalo:2020nvw} and the Liquid Scintillator Neutrino Detector (LSND)~(see~\cite{Aguilar:2001ty}, and references therein), have observed electron-like event excesses which have withstood scrutiny so far and which  cannot be easily explained within the ambit of the SM. We summarize, in turn,  the experiments,  their results, and the efforts to explain them. Prior to proceeding, we note that while the discussion in this  section covers both  LSND and  MB, given $a)$ the fact that both see electron-like excesses and $b)$ the many attempts to jointly explain them, our focus in the rest of the paper is the MB low-energy excess (LEE) and the anomalous magnetic moment of the muon. However, since the process chosen is, in principle, capable of giving the LSND final state, we also mention the implications for this in Section~\ref{sec4-A}, as well as discussing the consequences for the KARMEN experiment~\cite{EITEL200289}, which found a null result in its search for an LSND-like excess.

MB, at Fermilab,  uses  muon neutrino and anti-neutrino beams produced by 8~GeV protons hitting a beryllium target, with the fluxes peaking at around 600~MeV ($\nu_{\mu}$) and around 400~MeV ($\bar{\nu}_{\mu}$). The detector is a 40-foot diameter sphere containing  818 tons of pure mineral oil (CH$_2$) and is located 541~m from the target. Since 2002, the  MB experiment has collected a total of $11.27 \times 10^{20}$ POT in anti-neutrino mode and  $18.75 \times 10^{20}$ POT in neutrino mode. Quasi-elastic $e$-like event  excesses of $560.6 \pm 119.6$ in the neutrino mode, and $79.3 \pm 28.6$ in the anti-neutrino mode, with an overall significance of $4.8\sigma$ have been established in the neutrino energy range 200~MeV$< E^{QE}_{\nu} <$ 1250~MeV. In terms of visible energy, E$_{\rm vis}$, most of the excess is confined to the range $100$~MeV $<{\rm E}_{\rm vis}< 700$~MeV, with a somewhat forward angular distribution, and is referred to as the MB LEE. We note two points of relevance, $a)$ that all major backgrounds are constrained by \textit{in-situ} measurements, and $b)$ that  MB, being a mineral oil Cerenkov light detector, cannot distinguish photons from electrons in the final state. In addition,  MB,  under certain conditions (which we describe in more detail below) would also mis-identify an $e^+e^-$ pair as a single electron or positron.

LSND was  a detector  with 167 tons of mineral oil, lightly doped with scintillator. Neutrino and anti-neutrino beams originating from $\pi^-$ decay-in-flight (DIF) as well as $\mu$ decay-at-rest  (DAR) were used. The main interaction was the inverse beta decay process, $\anue + p \rightarrow e^+ + n$.  The final state  observed in the detector was the Cherenkov and scintillation light associated with the $e^+$ and the co-related and delayed scintillation light from the neutron capture on hydrogen, producing a 2.2~MeV $\gamma$. The experiment observed $87.9 \pm 22.4 \pm 6.0$ such events above expectations, at a significance of $3.8\sigma$, over its run span from 1993 to 1998 at the Los Alamos Accelerator National Laboratory. Like  MB,  LSND was unable to discriminate a photon signal from those of $e^+$, $e^-$ or an $e^+e^-$ pair.
\vspace{-1cm}
\subsection{Sterile neutrinos and other proposed new physics solutions of the MB and  LSND anomalies}

Perhaps the most widely discussed resolution of the MB and  LSND  excesses involves the presence of sterile neutrinos with mass-squared values of $\sim 1$~eV$^2$, mixed  with the SM neutrinos, leading to oscillations and $\anue$ and $\nue$ appearance~\cite{Aguilar-Arevalo:2018gpe}. Support to the sterile hypothesis is lent by deficits in $\nu_e$ events in radioactive source experiments~\cite{0711.4222,1006.3244} and in $\bar{\nu}_e$  reactor flux measurements~\cite{1101.2663,1101.2755,1106.0687,1309.4146,1605.02047}. Recent results from the reactor experiments,  NEOS~\cite{1610.05134} and DANSS~\cite{1606.02896} also provide hints of oscillations involving sterile neutrinos. As  other disappearance oscillation data sets and null results from multiple experiments have accumulated, however, this explanation for  MB and  LSND excesses has been subject to strongly increasing tension with their conclusions. In particular, results from MINOS/MINOS$+$~\cite{1710.06488} and IceCube~\cite{1605.01990} disappearance measurements constrain $\numu$ mixing with a sterile neutrino very strongly, in conflict with the demands of the  appearance hypothesis for  MB and  LSND. For recent global analyses and more detailed discussions, the reader is referred to~\cite{1204.5379,1602.00671,1607.00011, 1609.07803, 1703.00860, Dentler:2018sju, 1906.00045}. Finally, the presence of a light sterile neutrino is also disfavoured by cosmological data~\cite{1505.01076,1502.01589}.

This growing tension and the tightening of constraints on the presence of sterile neutrinos has led to efforts to find non-oscillatory solutions to one or both of these excesses. Earlier  attempts~\cite{hep-ph/0505216, 0902.3802,1009.5536,1210.1519} have typically included a heavy sterile ($\it{i.e.}$  dark) decaying neutrino which mixes with the SM active neutrinos. In proposals where the decay of the heavy neutrino is radiative~\cite{0902.3802,1009.5536}, there appears to be some conflict with  either tight constraints on mixings and magnetic moments~\cite{Bolton:2019pcu,0901.3589,1011.3046,1110.1610,1502.00477,1511.00683,1904.06787,1909.11198} or matching~\cite{1210.1519} the observed angular distribution of the visible light in  MB. Other efforts invoking  new physics include~\cite{1512.05357,1602.08766,1708.09548,1712.08019}, which appear to be  in tension with the conclusions of global analyses~\cite{1602.00671,1607.00011,1703.00860,Dentler:2018sju}. Among more recent work we list~\cite{1807.09877,1808.02915}, which involve the production and fast decay of a heavy neutrino in  MB, resulting in a collimated $e^+e^-$ pair;~\cite{1909.08571} which depends on an altered ratio of single photon to $\pi^0$ events  and~\cite{Fischer:2019fbw} which invokes the production  of a heavy neutrino in kaon decays in the proton beam target and its subsequent radiative decay.  There have  also been proposals~\cite{Dentler:2019dhz,deGouvea:2019qre}  which extend  the decay scenario proposed in~\cite{hep-ph/0505216}, originally proposed to explain  LSND, and apply it to MB. Most recently~\cite{Datta:2020auq,Dutta:2020scq} discuss  scalar mediated scenarios  which also address the KOTO and the  $g-2$ anomalies in addition to MB, while~\cite{Abdullahi:2020nyr} discusses it as well as a possible solution to $g-2$ and the  BaBar monophoton excess.
\vspace{-0.5cm}
\subsection{General constraint considerations relevant to new physics proposals for the MB LEE}
While we discuss the constraints on our specific model in more detail later in this work, we list here some that are particularly important to most efforts to explain the MB LEE. Any explanation involving the production of dark sector particles in the target  which then scatter elastically off the nucleons or electrons in the MB detector must confront the MB DM search results~\cite{1807.06137} which found no excess events in the off-target, \textit{i.e.} beam dump mode. This result signals that when neutrino production was suppressed via charged pion absorption in the beam dump (\textit{i.e.}, the target was removed) the excess disappeared.  Another class of important constraints are those arising from neutrino-electron scattering measurements~\cite{0101039,0911.1597,1104.1816,Park:2013dax,VILAIN1994246,1502.07763}. For a discussions of these constraints in the context of the MB LEE, see~\cite{Arguelles:2018mtc}.  Finally, as we show below, a set of constraints important to any new physics proposal that involves a new coupling to baryons and a direct or indirect coupling to neutrinos  originate from observations of neutral current neutrino-nucleon scattering at both low and high energies. At low energies, such a proposal must confront measurements such as those carried out by  MB~\cite{0909.4617}. At high energies, the deep inelastic neutrino-nucleon cross sections~\cite{ hep-ph/9512364, hep-ph/9807264, 1106.3723} are well understood and tested by HERA data~\cite{Chekanov:2002pv} all the way up to neutrino energies of $10^7$~GeV, and these results must be complied with. Finally, a recent general treatment focussed on the MB LEE which brings out the difficulties and constraints associated with finding a solution to this anomaly may be found in~\cite{Brdar:2020tle}.
\subsection{The muon and electron $g-2$ anomalies}
The Lande $g$ factor, and its deviation from the tree level value of $2$, is one of the most precisely measured quantities in the SM. This also renders it an excellent probe for new physics. At the present time, there exists a long-standing and statistically significant discrepancy between its measurement~\cite{Bennett:2006fi,Brown:2001mga} and the theoretically predicted value, which involves contributions from quantum electrodynamics, quantum chromodynamics and electroweak theory~\cite{Miller:2007kk,  Jegerlehner:2009ry,Lindner:2016bgg, Holzbauer:2016cnd,Davier:2019can,Aoyama:2020ynm}. Specifically, a $3.7\sigma$ muon $g-2$ discrepancy has been found as follows~\cite{Aoyama:2020ynm}
\begin{equation}
\Delta a_\mu = a_\mu^{\rm meas}-a_\mu^{\rm theory}=(2.79\pm 0.76)\times 10^{-9}.
\end{equation}
Many proposals for new physics provide possible explanations for this discrepancy (For reviews and references, see~\cite{Miller:2007kk,Jegerlehner:2009ry,Lindner:2016bgg, Holzbauer:2016cnd}.). Our attempt in this work, details of which are provided in the sections to follow, is related to a class of possible solutions suggested by several authors~\cite{Kinoshita:1990aj,Zhou:2001ew, Barger:2010aj, TuckerSmith:2010ra, Chen:2015vqy, Liu:2016qwd, Batell:2016ove, Marciano:2016yhf, Wang:2016ggf, Liu:2018xkx, Liu:2020qgx, Jana:2020pxx} involving a light scalar with a mass in the sub-GeV range and a relatively weak coupling to muons.

Also, from the high precision
measurement of the fine structure constant, a $2.4\sigma$ discrepancy has been recently found between the theoretical value and experimental measurement of the electron magnetic moment~\cite{Parker:2018vye},
\begin{equation}
\Delta a_e = a_e^{\rm exp}-a_e^{\rm theory}=(-8.7\pm 3.6)\times 10^{-13}.
\end{equation}
\vspace{-0.5cm}
\section{The Model, its motivations and the interaction in MiniBooNE}
\label{sec3}
\subsection{Motivations for the choice of the additional $U(1)$}
For reasons enumerated in the beginning of the previous section, one may legitimately assume that the SM is a highly successful  low energy effective description of a more fundamental and complete theory. Effective field theories are not, in general, expected to satisfy the stringent requirements of renormalizability and anomaly cancellation, and yet the SM does satisfy these important criteria. One may choose to treat this as a curious accident, or one could adopt it as a guiding principle and impose ultra-violet (UV) completion and  the freedom from anomalies as a desirable requirement~\cite{1812.04602} when considering a further $U(1)$ extension. We choose this approach for arriving at one of the benchmark choices we make here ($U(1)_{B-3L_\tau}$, below).  From a phenomenological point of view, however, we find that a second option which does not satisfy these criteria, a $U(1)$ with gauged baryon number, works equally well for explaining the MB LEE and accommodating the muon and electron~$g-2$. This latter choice must, however, be supplemented by a set of heavy chiral fermions.

The  global symmetries of the SM, namely, $U(1)_B, U(1)_{L_e}, U(1)_{L_\mu}$ and $U(1)_{L_\tau}$, provide possible signposts to an extension. These lead to three combinations which are anomaly-free and consequently do not require the addition of any new fermions, {\it i.e.}, $  U(1)_{L_{\mu}-L_e}, U(1)_{L_e-L_\tau}$ and $U(1)_{L_\mu-L_\tau}$~\cite{Foot:1990mn,PhysRevD.43.R22,PhysRevD.44.2118}. In addition,  if right-handed (RH) neutrinos are added to the SM particle spectrum,  it can be shown~\cite{1203.4951, 1812.04602, Heeck:2018nzc} that  $U(1)_{B-L} \times U(1)_{L_\mu-L_\tau}\times U(1)_{L_\mu-L_e}$ or any of its subgroups provide anomaly-free and UV complete options for adding a new  $U(1) $ gauge boson to the SM. Noting that $a)$ the necessary new physics to explain  MB  must couple  neutrinos  to baryons either directly or via mixing, (since the incoming beam is a $\numu$ or a $\anumu$ and  the target nucleus is CH$_2$) and, $b)$ that  a universal coupling to the quark generations ensures safety from flavor changing neutral  currents (FCNCs), one is led to a  class of symmetries, $\ie$  $B- r_{\ell} L_{\ell}$, with $r_{\ell} L_{\ell}  = 3,$ where the $r_{\ell}$ are real coefficients and $\ell=e,\mu,\tau$.

For several examples of  this general class of possibilities,  the phenomenology of and constraints on  the associated  boson have been studied in~\cite{1210.2703,1408.6845,1502.07763,1512.03179,1801.04847,1803.05466}. They arise from beam dump, fixed target, collider, weak precision and neutrino  experiments (for a complete list, see~\cite{1803.05466}  and references therein) which tightly restrict  the gauge coupling and the mass of the new gauge boson. Additional constraints on electron couplings arise from neutrino electron elastic scattering experiments~\cite{0911.1597,1803.01224,VILAIN1994246,Park:2013dax}. Overall, one is led to the conclusion that it is very difficult to explain the MB LEE and simultaneously satisfy all constraints on a $U(1)$ if it couples to any significant degree to electrons. Based on this,  possibilities, like $U(1)_{B-3/2(L_\mu + L_\tau)}, U(1)_{B-3L_\mu}$, and $U(1)_{B-3L_\tau}$, which, while also tightly constrained~\cite{Heeck:2018nzc,Han:2019zkz}, offer a little more room for accommodating new physics explanations. In our work, we have chosen to use $U(1)_{B-3L_\tau}$~\cite{Ma:1997nq, Ma:1998dp, Ma:1998dr} as an example, but it is possible that a fuller exploration of the possibilities available may yield other equivalent anomaly-free and UV complete options among the larger set  $B- r_{\ell} L_{\ell}$ identified above. 

As mentioned, $U(1)_B$ affords a phenomenologically equivalent alternative insofar as explaining the two anomalous results we focus on in our work. Gauging baryon number alone has been discussed extensively in the literature~\cite{NELSON198980,PhysRevD.18.242,He:1989mi,PhysRevD.40.2487,PhysRevLett.74.3122,Carone:1995pu,1002.1754,1005.0617,1010.3818,1012.4679,1104.3145,1105.3190,1106.0343,1106.4347, 1107.2666,1203.0545,1404.3947,1404.4370,1810.06646}. A gauged $U(1)_B$, unlike the accidental SM symmetry combinations mentioned above, is not anomaly-free and must be treated as an effective theory with an UV cut-off, with new states entering at higher energies to make the theory consistent. A discussion of the necessary UV completions is outside the scope of our work and we refer the reader to the references above for examples of such models.
\subsection{Some other considerations}
The associated gauge boson ($Z'$) for both our example gauge groups also couples to the dark sector. We note that  there are observational reasons that hint towards a link that may exist between DM and baryons. These are  the stability of both DM and protons on a timescale equal to or exceeding the age of the universe, and the empirically known but unexplained fact that the relic abundances of baryons are similar to those of DM up to a factor of $\sim 5$~\cite{1107.2666}. The $Z'$ in our work is a portal particle, coupled via  $U(1)_{B-3L_\tau}$ (or  $U(1)_{B}$) to the SM with a coupling $g_B$ and to the dark sector via a coupling $g_d$.

Prior to providing details of the model and the interaction in  MB in the two next sections, we discuss two important gauge invariant and renormalizeable  terms associated with any new $U(1)$ that is linked to  the SM, specifically to its $U(1)_Y$ hypercharge group. These involve kinetic~\cite{Holdom:1985ag} and mass mixings.  After convenient field redefinitions (see, {\it e.g.},~\cite{1203.2947}) they enter the Lagrangian as
\begin{equation}
{\cal L} \supset  e~\epsilon \, Z'^\mu \, 
J_\mu^{\rm em}
+ \frac{g}{c_W} \epsilon' \, Z'^\mu \, 
J_\mu^{\rm Z} \,,
\label{eq:mix}
\end{equation}
where $Z$ and $Z'$  are the weak neutral SM and new gauge bosons, $J_\mu^{\rm em}$ and $J_\mu^{\rm Z}$ the electromagnetic and $Z$ currents, $e$ is the usual electric charge, $g$ is the weak gauge coupling, $c_W$ is the cosine of the Weinberg angle, and $\epsilon$ and $\epsilon'$  parameterize the kinetic and mass mixings, respectively. In situations where $\epsilon$ ($\epsilon'$) is sizeable and has measurable phenomenological consequences for current or near-future experiments, the  $Z'$ is usually referred to as a ``dark photon" (``dark $Z$"). In general, even if one assumes that kinetic mixing vanishes at high energies, it re-appears via loop effects. Specifically,  if there are $i$ particles with mass $M_i$ charged under both $U(1)_Y$ and the new $U(1)_{Z'}$ with couplings $g_Y$ and $g_{Z'}$ respectively, kinetic mixing is generated at the loop level with a magnitude~\cite{Holdom:1985ag, Cheung:2009qd}
\begin{equation}
\epsilon=\frac{{g_Y} g_{Z'}}{16\pi^2}\sum_{i}{{q^i_Y q^i_{Z'}} \ln{\frac{M_i^2}{\mu^2}}},
\end{equation}
where $q_Y$ and $q_{Z'}$ are the respective charges and $\mu$ is a renormalization scale. In what follows, $g_{Z'} = g_{B-3L_\tau}$ or  $g_{Z'} = g_{B}$ is constrained to be $\simeq 10^{-4}$ or smaller (see Fig.~\ref{MZp-constraints-fig}), rendering $\epsilon$ very small. This allows us to assume in what follows that the main decay modes of the $Z'$ are to invisible particles of the dark sector.  Finally, we note that  kinetic  mixing may also be naturally small  below the electroweak scale, if in the full theory at high energy the $U(1)_{Z'}$ is actually embedded in a larger non-abelian gauge group~\cite{Carone:1995pu}.

The mass mixing $\epsilon'$ between the $Z$ and $Z'$ at tree level arises if there is a scalar which acquires a vacuum expectation value (vev) and is charged under both the $U(1)_Y$ and the $U(1)_{Z'}$. Given the fact that our model does not contain such a particle, and that $\epsilon'\propto m_{Z'}/m_Z$ \cite{1203.2947} which is quite small, we also neglect the mass mixing term proportional to $\epsilon'$ in addition to $\epsilon$.  For completeness, we mention that constraints on the kinetic mixing of dark photons for low mass $Z'$ are very severe, and arise from a large number of collider, neutrino, beam dump and other experiments; for a recent comprehensive discussion and list of references the reader is referred to~\cite{Essig:2013lka,1801.04847}. The physics of and constraints on a mass-mixed $Z'$ are discussed in~\cite{1203.2947,1205.2709,Essig:2013lka}.

Finally, we extend the scalar sector of the SM by adding a second Higgs doublet, $\ie$, the widely studied two Higgs doublet model (2HDM)~\cite{Lee:1973iz, Branco:2011iw} and add $i)$ a dark sector singlet scalar $\phi_{h'}$  which acquires a vev and gives mass to the $Z'$, and $ii)$ a dark neutrino $\nu_d$. The process we consider in order to explain the MB LEE involves a beam $\numu$, which produces, (via mixing)  a dark neutrino ($\nu_4$), which is the  mass eigenstate corresponding to $\nu_d$. Also present in the final state are   $i)$ a recoiling nucleon (incoherent scattering) or  nucleus (coherent scattering) and $ii)$ a light  scalar $h'$ or $H$,  which quickly decays to an $e^+e^-$ pair.  The scattering is mediated by  $Z'$, as shown in (Fig.~\ref{FD-SP-MB}). RH neutrinos are introduced for the purpose of anomaly cancellation and for generating neutrino masses via the seesaw mechanism. Further details are provided in the sections below.
\subsection{The Lagrangian of the model}
As discussed above,  the  SM is extended  by a second Higgs doublet, and either $i)$ a $U(1)_{B-3L_\tau}$ gauge boson, coupled to baryons and  the $\tau$ sector  or $ii)$ a $U(1)_B$ gauge boson coupled to baryon number alone with gauge coupling $g_B$\footnote{In the remainder of our work, we use $g_B$ as a generic notation for both the $U(1)_B$ coupling and/or the $U(1)_{B-3L_\tau}$ coupling  for the most part, specifying $g_{B-3L_\tau}$ only when the context demands it, as, for instance, in Fig.~\ref{MZp-constraints-fig} and Section~\ref{sec5}. We stress that in the numerical calculations, they correspond to the same values.}, with no tree level couplings to the leptons of the SM. In both cases the coupling to the incoming muon neutrinos is indirectly generated via mixing with the dark neutrino $\nu_d$, since the light new mediator $Z'$  couples to it  with coupling $g_d$. As may be seen from Table~\ref{tab2}, which lists the benchmark values we use below, we have assumed $$ g_B \ll g_d,$$  essentially dictated by constraints that we discuss in Section~\ref{sec5-A}. Such a hierarchy of couplings could effectively arise, of course, from widely differing charges for the same gauge boson. Perhaps a more natural possibility~\cite{Lebedev:2014bba} is to assume that the disparity originates in the mixing of two $U(1)$ gauge bosons $Z_1$ and $Z_2$, with significantly different  mass eigenvalues  $m_1 \ll m_2$,  with $Z_1$ coupling to only the dark sector  and $Z_2$ coupling only to  SM particles. The lighter mass eigenstate, a mixture of  $Z_1$ and $Z_2$, would then be effectively coupled to the SM with a coupling $g_B \sim g_d\, m_1^2/m_2^2$. A second possibility~\cite{Fox:2011qd}  leading to a $g_B\ll g_d$ involves an effective $Z'$, which has no tree level SM couplings but couples via non-renormalizable operators. 

The SM Lagrangian is thus extended by the following terms to obtain $\mathcal{L}_{\rm tot}$, the full Lagrangian of the extended theory,

\begin{equation}
{\cal L}_{\rm tot}\supset -\frac{1}{4} Z'_{\mu\nu}Z'^{\mu\nu}+\bar{\nu}_d\gamma^\mu(i\partial_\mu+g_d Z'_\mu)\nu_d+{\cal L}_q+{\cal L}_f-{\cal L}_Y^f-V+{\cal L}_S^{\rm Kin}+{\cal L}_m,
\end{equation}
where
\begin{eqnarray}
&{\cal L}_q=\sum_{q}{\frac{1}{3}g_B \bar q\gamma^\mu Z'_\mu q},~~~{\cal L}_{f} =\sum_{f}{g_B q_f \bar{f} \gamma^\mu Z'_\mu f},\\
&{\cal L}_{Y}^f\!=\!\!\sqrt{2}\Big[{(Y^u_{ij} \tilde\Phi_1 +\tilde{Y}^u_{ij} \tilde\Phi_2) \bar{Q}_L^i  u^j_R}+{(Y^d_{ij}  \Phi_1 +\tilde{Y}^d_{ij} \Phi_2) \bar{Q}_L^i  d^j_R}+{(Y^e_{ij}  \Phi_1 +\tilde{Y}^e_{ij} \Phi_2) \bar{L}_L^i  e^j_R}+h.c.\Big]\!.\label{Eq:LYe}
\end{eqnarray}
{In the above, $q$ runs over all the SM quarks, while $f$ runs over the leptons with charge $q_f$ to which $Z'$ is coupled to, $\eg$ $\nu_\tau$, $\tau$ for our choice of $U(1)_{B-3L_\tau}$, and over none of the lepton generations for $U(1)_B$. In Eq.~(\ref{Eq:LYe}), $Q_L,u_R,d_R$ are the left-handed (LH) quark doublets, RH up-type quarks and RH down-type quarks respectively. Similarly, $L_L$ and $e_R$ denote the LH SM lepton doublets and the RH charged leptons, respectively. $\Phi_1$ and $\Phi_2$ are the two doublets of the 2HDM, and $Y_{ij}$ and $\tilde{Y}_{ij}$ are the associated Yukawa coupling matrices.

Our approach with respect to the 2HDM in this section is similar to that followed in~\cite{Jana:2020pxx}. We  write  the scalar potential $V$ in the Higgs basis $(\phi_h,\phi_H,\phi_{h'})$~\cite{Branco:1999fs,Davidson:2005cw}, with the $\lambda_i$ denoting the usual set of quartic couplings}
\begin{eqnarray}
V &=& \phi_{h}^\dagger\phi_{h}\left(\frac{\lambda_1}{2} \phi_{h}^\dagger\phi_{h}+\lambda_3 \phi_{H}^\dagger\phi_{H}+\mu_1\right) +\phi_{H}^\dagger\phi_{H}\left(\frac{\lambda_2}{2} \phi_{H}^\dagger\phi_{H}+\mu_2\right)+\lambda_4 (\phi_{h}^\dagger\phi_{H})(\phi_{H}^\dagger\phi_{h})\nonumber\\
&+&\left\{\left(\frac{\lambda_5}{2} \phi_{h}^\dagger\phi_{H} +\lambda_6\phi_{h}^\dagger\phi_{h}+ \lambda_7  \phi_{H}^\dagger\phi_{H}+ \lambda'_5 \phi_{h'}^\ast\phi_{h'}- \mu_{12}\right)\phi_{h}^\dagger\phi_{H}+h.c.\right\}\nonumber\\
&+&\phi_{h'}^\ast\phi_{h'}(\lambda'_2 \phi_{h'}^\ast\phi_{h'}+\lambda'_3 \phi_{h}^\dagger\phi_{h}+\lambda'_4\phi_{H}^\dagger\phi_{H}+\mu'),
\end{eqnarray}
where
\begin{eqnarray}
\phi_{h}=\left( \begin{array}{c}
H^+_1 \\
\frac{v+H^0_1+i G^0_1}{\sqrt{2}} \\
\end{array} \right)\equiv\cos\beta\, \Phi_1+\sin\beta\,\Phi_2
,~
\phi_{H}=\left( \begin{array}{c}
H^+_2 \\
\frac{H^0_2+i A^0}{\sqrt{2}} \\
\end{array} \right)\equiv-\sin\beta\, \Phi_1+\cos\beta\,\Phi_2,
\end{eqnarray}
\begin{equation}
\phi_{h'}=\frac{v'+H^0_3+i G^0_2}{\sqrt{2}},
\end{equation}
so that $v^2 =v^2_1+v^2_2 \simeq  (246~{\rm GeV})^2$ and $\tan\beta=v_2/v_1$, where $\langle \Phi_i \rangle = v_i/\sqrt{2}$ and $\langle\phi_{h'}\rangle=v'/\sqrt{2}$. Here, $H_1^+,G^0_i$ are the Goldstone bosons eaten up by the gauge bosons after the electroweak and $U(1)'$ symmetries are spontaneously broken. Therefore, the scalar kinetic term ${\cal L}_S^{\rm Kin}$ can be written as
\begin{equation}
{\cal L}_S^{\rm Kin}=\sum_{\cal H}(D_\mu^{\cal H}\phi_{\cal H})^\dagger D_\mu^{\cal H}\phi_{\cal H} \supset \frac{1}{2} g_d^2(v'+H^0_3)^2 Z'_\mu Z'^\mu,
\end{equation}
where
\begin{equation}
D_\mu^{h'} \phi_{h'}\equiv (\partial_\mu + i g_d Z'_\mu)\phi_{h'}.
\end{equation}
Hence, the $Z'$-$H_3^0$-$Z'$ coupling is given by
\begin{equation}
G_{Z'Z'H_3^0}=i\frac{2 m_{Z'}^2}{v'},
\end{equation}  
where $~ m_{Z'}^2=g_d^2 v'^2$.
The mass matrix of the neutral CP-even Higgses in the basis:
$\left(H_1^0, H_2^0, H_3^0\right)$ is given by
\begin{equation} 
m^2_{\cal H} = \left( 
\begin{array}{ccc}
\lambda_1 v^2 &\lambda_6 v^2  &  \lambda'_3 v v'\\ 
\lambda_6 v^2 & \bar{m}_H^2 & \lambda'_5 v v' \\ 
 \lambda'_3 v v' & \lambda'_5 v v' &2 \lambda'_2 v'^2\end{array} 
\right), 
 \end{equation} 
where $\bar{m}_H^2 = \mu_2 +(\lambda_3 + \lambda_4 + \lambda_5)v^2/2 + \lambda'_4
v'^2/2$. Here, we have used the following minimization conditions of the scalar potential $V$,
\begin{eqnarray}
\mu_1 &=& -\frac{1}{2}(\lambda_1 v^2+\lambda'_3 v'^2),\\
\mu_{12} &=& \frac{1}{2}(\lambda_6 v^2+\lambda'_5 v'^2),\\
\mu' &=& -\lambda'_2 v'^2-\frac{\lambda'_3 v^2}{2}.
\end{eqnarray} 
The mass matrix of the neutral CP-even Higgses $m^2_{\cal H}$ is diagonalized by \(Z^{\cal H}\) as follows (see appendix~\ref{App:AppendixA}): 
\begin{equation} 
Z^{\cal H} m^2_{\cal H} (Z^{\cal H})^T = (m^{2}_{\cal H})^{\rm diag}\,, 
~~~~{\rm with}~~~~
H^0_i = \sum_{j}Z_{{j i}}^{\cal H}h_{{j}}\,, 
\end{equation} 
where $(h_1,h_2,h_3)=(h,H,h')$ are the mass eigenstates, and  $H^0_1\approx h$ is the SM-like Higgs in the alignment limit (\ie, $\lambda_6\sim0\sim\lambda'_3$) assumed here. The  masses of the CP-even physical Higgs states $(h,H,h')$ are given by
\begin{equation}
m^{2}_{h,H,h'}\simeq\left\{\lambda_1 v^2,\frac{1}{2}\left(\bar{m}_H^2+2 \lambda'_2 v'^2\pm \sqrt{(\bar{m}_H^2-2
\lambda'_2 v'^2)^2 + 4 (\lambda'_5 v v')^2 }\right)\right\}.
\end{equation}
Also, in the present model, the charged and CP-odd Higgs masses, respectively, are given by 
\begin{eqnarray}
m^2_{H^\pm}&=&\frac{1}{2} \left(2 \mu_2+\lambda_3 v^2+\lambda_4' v'^2\right),\\
m^2_A&=&\frac{1}{2} \left(2 \mu _2+\left(\lambda _3+\lambda _4-\lambda _5\right) v^2+\lambda_4' v'^2\right).
\end{eqnarray}
Our explanation of the muon and electron $g-2$ draws upon contributions from two light scalars, $h'$ and $H$, leading to  $m^2_{H},m^2_{h'} \ll m^2_{H^\pm},  m^2_{A}$. As discussed in a later section, electroweak precision measurements, expressed in terms of oblique parameters, lead to a mass hierarchy $m_A\sim m_{H^\pm}$ $\gg m_H$. In addition, collider constraints (discussed in a later section below) set a lower bound on $m_{H^\pm}$, requiring it to be comfortably above  $\sim 110$ GeV. For our purpose, we assume  $m^2_{H^{\pm}}\simeq m^2_A$. They do not play an essential role in our scenario, and we have checked that  contributions made by them to the muon and electron $g-2$ are negligibly small. The necessary closeness in mass then implies  $\lambda_4\simeq\lambda_5$, leading to $m^2_{H^\pm}=m^2_A\simeq-v^2 \lambda_5$. Perturbativity ($|\lambda_5|\lesssim \sqrt{4\pi}$) then imposes an upper bound on these masses,  $m_{H^\pm}=m_A \lesssim 460$~GeV, with $m_H$ thus restricted to be $\sim$~GeV or less~\cite{Jana:2020pxx}.

As we discuss in a later section, LEP allows us to obtain a lower bound  on the charged Higgs, $\ie$ $m_{H^\pm}\simeq v \sqrt{|\lambda_5|}\geq 110$~GeV. This upper bound can be then translated to $|\lambda_5|\geq 0.2$. This is relatively insensitive to mass in the low mass region,$ \ie$  $m_{H} \leq 1$~GeV. 

In the Higgs basis the Lagrangian ${\cal L}_Y^f$ can be written as follows
\begin{eqnarray}
{\cal L}_{Y}^f \!&=&\!\sqrt{2}\Big[(X^u_{ij} \tilde\phi_h\!\! +\!\!\bar{X}^u_{ij} \tilde\phi_H) \bar{Q}_L^i  u^j_R\!+\!(X^d_{ij}  \phi_h \!\!+\!\!\bar{X}^d_{ij} \phi_H) \bar{Q}_L^i  d^j_R\!+\!(X^e_{ij}  \phi_h \!\!+\!\!\bar{X}^e_{ij} \phi_H) \bar{L}_L^i  e^j_R\!+\!h.c.\Big]\!,
\end{eqnarray}
where
\begin{eqnarray}
X^k_{ij}&=& Y^k_{ij} \cos\beta +\tilde{Y}^k_{ij} \sin\beta,\\
\bar{X}^k_{ij}&=& - Y^k_{ij} \sin\beta +\tilde{Y}^k_{ij} \cos\beta.
\end{eqnarray}
We emphasize that $X^k_{ij}$ and $\bar{X}^k_{ij}$ are independent Yukawa matrices. Moreover, the fermion masses receive contributions only  from $X^k_{ij}$, since  in the Higgs basis only $\phi_h$ acquires a non-zero vev. This leads to  $X^k= {\cal M}_k/v$, where ${\cal M}_k$ are the fermion mass matrices.   Hereafter, we work in a basis in which the fermion mass matrices are real and diagonal, where $U_k {\cal M}_k V^\dagger_k=m_k^{\rm diag}$ are their bi-unitary transformations. In this basis, in general, $\bar X^k_{ij}$ are free parameters and non-diagonal matrices.

From the leptonic Lagrangian ${\cal L}_Y^\ell$, their  interactions with the physical scalar states are given by
\begin{equation}
{\cal L}_Y^\ell=\sum_{\ell=e, \mu, \tau}[{X}^\ell_{ij}h +  \bar{X}^\ell_{ij} (Z_{32}^{\cal H}h'+ Z_{22}^{\cal H} H)] \bar{\ell}_L^i  \ell^j_R+h.c.,
\end{equation}
one finds the following coupling strengths of the scalars $h, h',H$ with a lepton pair, respectively: 
\begin{equation}\label{phi-ll-coupling}
y^{h}_{\ell} 
=\frac{m_\ell}{v},~~~~y^{h'}_{\ell} 
=y^\ell Z_{{3 2}}^{\cal H}=y^\ell \sin\delta,~~~~y^{H}_{\ell} 
=y^\ell Z_{{2 2}}^{\cal H}=y^\ell\cos\delta,
\end{equation}
where ${\rm diag}\{m_e,m_\mu,m_\tau\}=U_\ell {\cal M}_\ell V^\dagger_\ell$ and $\delta$ is the scalar mixing angle between the mass eigenstates $(H,h')$ and the gauge eigenstates ($H^0_2, H^0_3$).  In the above, we work in the mass basis where the diagonal elements of the rotated $\bar{X}^\ell_{ij}={\rm diag}\{y^e,y^\mu,y^\tau\}$ and to avoid most of flavor violating processes and explain electron $g-2$ simultaneously, we have chosen all off-diagonal elements to be zero except $y^{e\tau}$ and $y^{\tau e}$, as we discuss in Section~\ref{sec4-C}. Additionally, the quark $\bar{X}^k_{ij}$ are assumed to be very small to suppress flavor violating processes. An example of an ansatz that can achieve such suppression is discussed in~\cite{Babu:2018uik}.

Finally, ${\cal L}_m$ represents mass terms for the SM fermions, weak gauge bosons and the neutrinos. The full neutrino  mass matrix contains mass terms for both the SM neutrinos and the additional ones we introduce, since the masses are linked to each other at the Lagrangian level. In addition to a LH $\nu_d$, we have a RH partner ($N^4_R$) to cancel the $[ U(1)']^3$ anomaly in the dark sector, as well as three RH neutrinos ($N^i_R, i =1,2,3$) to achieve the usual SM anomaly cancellation. We further assume that the mass eigenstates of the RH neutrinos are large (to induce the see-saw mechanism) and can be integrated out. Thus at the low energies of interest to us here, one is left with a $4\times4$ mixing matrix $U$, which connects the flavor states $e,\mu,\tau,d$ to the mass eigenstates $\nu_1, \nu_2, \nu_3$ and $\nu_4$, and $U_{\rm PMNS}$, the usual SM lepton sector mixing matrix is a $3\times3$ sub-matrix of $U$.
\subsection{The interaction in MiniBooNE}
\begin{figure}[t!]
\center
\includegraphics[width=0.6\textwidth]{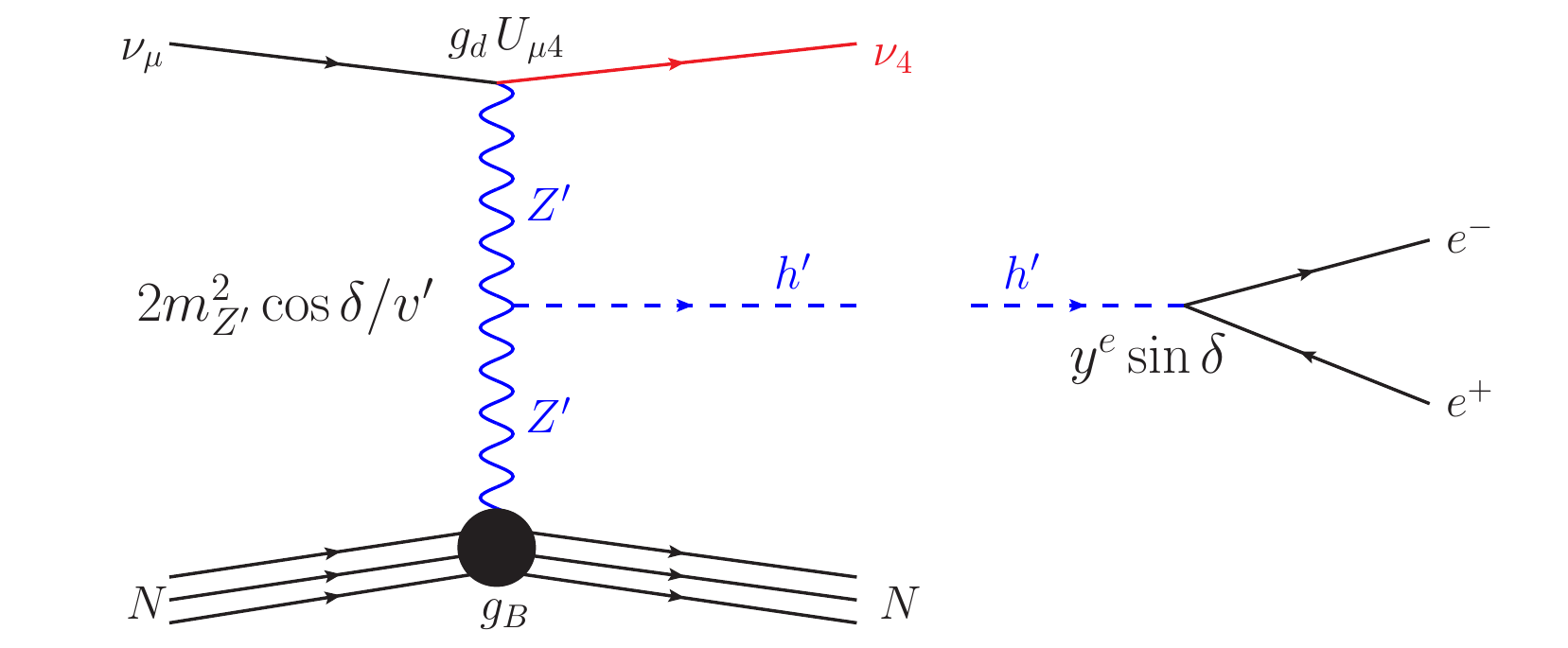}
\caption{Feynman diagram of the  scattering process in our model which leads to the excess in MB. Note that $H$ also contributes via the same diagram.}
\label{FD-SP-MB}
\end{figure}

As mentioned above, the dark neutrino ($\nu_d$) mixes with the standard massive neutrinos. Writing the interaction term in  the mass basis, we have
\begin{eqnarray}
\mathcal{L}_{\rm int} = - g_d \sum^{4}_{i,j=1} U^{\ast}_{di} U_{dj} \bar{\nu}_{i}\gamma^{\mu}Z'_{\mu}\dfrac{(1-\gamma_5)}{2} \nu_j.
\end{eqnarray}
The assumed value of the mass of the $\nu_4$ plays a somewhat secondary role in our calculation, and we comment here on its dependence, which arises primarily from kinematic considerations. Varying the mass within a range allowed by existing constraints does not affect the results in a qualitative manner. The benchmark value (see Table~\ref{tab2}) for its mass assumed in what follows  is $\sim$~50~MeV,  hence it will not be  produced in  pion decay. Thus, in our model, the MB  beam  primarily consists of $\nu_{\mu}$ produced via pion decay as  the superposition of  three mass eigenstates. The relevant process leading to an excess proceeds via  the new $Z'$, producing a collimated $e^+ e^-$ pair via the light scalar ($h'$) decay.  As part of  the final state, a $\nu_4$ kinematically accessible for the MB neutrino beam energy is also produced, making it  proportional to $|U_{\mu 4}|^2$,  as shown in the Fig.~\ref{FD-SP-MB}. In what follows, we have assumed that the $\nu_4$ does not decay visibly in MB after production. The  $Z^\prime$ couples to quarks via its coupling to baryon number, and consequently to nucleons, denoted below by $N$.
The on-shell matrix elements of the new $Z'$ neutral currents take the form
\begin{eqnarray} \nonumber
\langle N(k')|J^{\mu}_{Z'}| N(k)\rangle &=& g_{B} \bar{u}(k')  \Gamma^{\mu}_{Z'} (k'-k)u(k),
\end{eqnarray}
where, $k$ and $k'$ are the initial and final nucleon momenta, and  
\begin{eqnarray}
\Gamma^{\mu}_{Z'}(q) = \gamma^{\mu} F^{1}_{V}(q^2)  + \dfrac{i}{\,2 \, m_{N}}\sigma^{\mu \nu} q_{\nu} \, F^{2}_{V}(q^2).
\end{eqnarray}
The isoscalar form factors  $F^{1}_{V}(q^2)$ and $ F^{2}_{V}(q^2)$ for the nucleon are given by~\cite{Hill:2010yb}
\begin{equation}
\dfrac{F^{1}_{V}(q^2)}{F_{D}(q^2)}  =  1 - \dfrac{q^2 (a_{p}+ a_{n})}{4m^{2}_{N}- q^2},~~~~~\frac{F^{2}_{V}(q^2)}{F_{D}(q^2)} =  \dfrac{4m^{2}_{N}(a_p + a_n)}{4m^{2}_{N}- q^2 },
\end{equation}
where $m_N=0.938$~GeV, $F_{D}(q^2) = (1-q^{2}/0.71~\rm{GeV^{2}})^{-2}$, $a_p \approx 1.79$ and $a_n  \approx -1.91$ are coefficients related to the  magnetic moments of the proton and neutron, respectively.

To compute the total differential cross section, we consider both the incoherent and coherent contributions in the production of $h'$, as shown in Fig.~\ref{FD-SP-MB}. The total differential cross section, for the target in  MB,  $\ie$, CH$_2$, is given by
\begin{eqnarray}
\left(\dfrac{d\sigma}{dE_{h'}}\right)_{{\rm CH_{2}}} =\underbrace{14 \times \left(\dfrac{d\sigma}{dE_{h'}}\right)}_{\textrm{\footnotesize{{incoherent}}}}+ \underbrace{ 144 \times exp(2b(k'-k)^{2})\left(\dfrac{d\sigma}{dE_{h'}}\right)}_{\textrm{\footnotesize{{coherent}}}}.
\label{tot_xsec}
\end{eqnarray}
For the incoherent process, we have multiplied the single nucleon cross section by the total number of the nucleons present in CH$_2$  \textit{i.e.}, 14. In the coherent process the entire  carbon nucleus (C$^{12}$) contributes in the process and the contribution is large when the momentum transfer is small, \textit{i.e.} $q^2 = (k'-k)^2 \sim 0$. As $q^2$ increases, the coherent contributions are reduced significantly. This   is implemented by the form factor $exp(2b(k'-k)^{2})$~\cite{Hill:2009ek}, where $b$ is a numerical parameter, which for C$^{12}$, has been chosen to be $25$~GeV$^{-2}$~\cite{PhysRevD.9.1389, Hill:2009ek}. 

We have used Eq.~(\ref{tot_xsec}) to calculate the total number of $h'$
produced in the final state. Once $h'$ is produced, it decays promptly to
an $e^+ e^-$ pair, its lifetime being decided by its coupling to electrons.
Neglecting the mass of the electron, the lifetime of $h'$ is given by
\begin{eqnarray}
\tau_{h'} = \dfrac{8 \pi}{(y^{h'}_{e})^2 \,  m_{h'}}.
\label{h-life-tme}
\end{eqnarray}
For our benchmark parameter values, the lifetime of $h'$ is $3.5\times
10^{-13}$ seconds. We note that  MB is not able to distinguish an $e^+ e^-$
pair from a single electron~\cite{1810.07185, Karagiorgi:2010zz} if
$m_{\rm track} < 30$~MeV, where
\begin{eqnarray}
m_{\rm track} \equiv \sqrt{2E_1 E_2 (1-\cos \theta_{12})}.
\label{track-def}
\end{eqnarray}
Here $E_1$ and $E_2$ are the track energies and $\theta_{12}$ is the angle
between two tracks.
Since we have chosen the mass of $h'$ to be 23~MeV, the $m_{\rm track}$ produced by $h'$
decay is always less than 30~MeV. Hence, the decay of $h'$ to an $e^+ e^-$
pair mimics the single electron charged current quasi-elastic (CCQE) signal in the detector. We note that  $H$ can also contribute to the MB signal, since it can  be produced in the final state and subsequently  decay promptly  to an  $e^+e^-$ pair.  If the opening  angle of the two electrons is less than $8^\circ$ or one of electrons has  energy less than $30$~MeV, it would add to the signal. We find that only a  fraction $(\sim 10-15\%)$ of the  total number of the $H$ produced satisfy these criteria. Further suppression are provided by kinematics, since its mass is higher than that of $h'$, and by  $\sin^2 \delta$. Hence, the contribution of $H$ to the MB events is  small. Additionally, we have checked  that the production of two $h'$s, two $H$s or $h'H$ via the quartic couplings to $Z'$ is suppressed compared to single  $h'$ production in the final state.

Our results are presented in the next section\footnote{\vspace{-0.15cm}The results we present have been computed using our own code, and checked subsequently by implementing  the present model in \texttt{SARAH}~\cite{Staub:2008uz, Staub:2013tta} and by  using {\tt MadGraph5\_aMC@NLO}~\cite{Alwall:2014hca}.\vspace{0.2cm}}.
\section{Results}
\label{sec4}
In this section we present the results of our numerical calculations,  using the   cross section for the process and the model described in Section~\ref{sec3}.
\subsection{Results for  MiniBooNE and implications for  LSND and   KARMEN}
\label{sec4-A}
\begin{figure}[t!]
\includegraphics[width=0.49\textwidth]{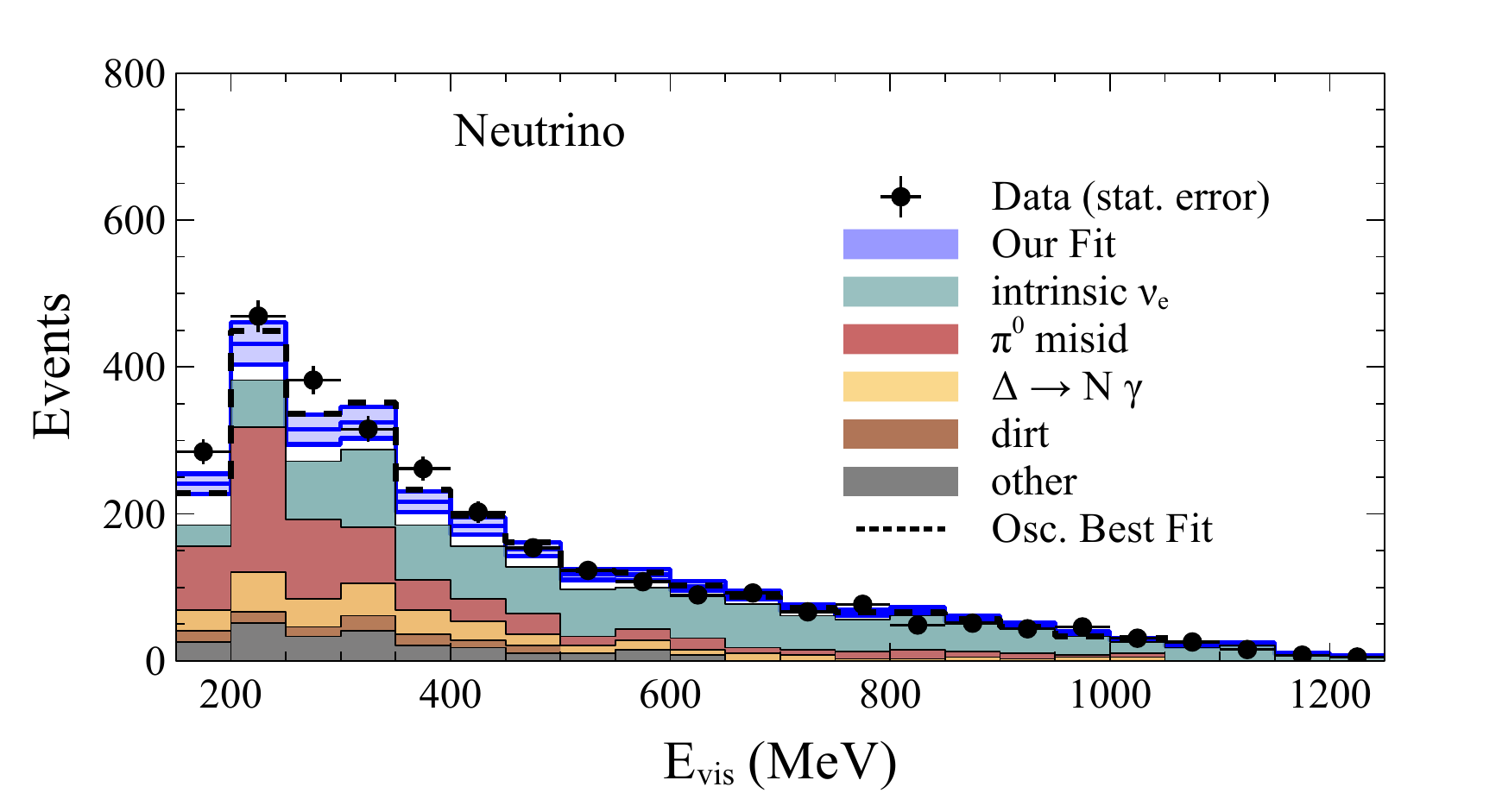}
\includegraphics[width=0.49\textwidth]{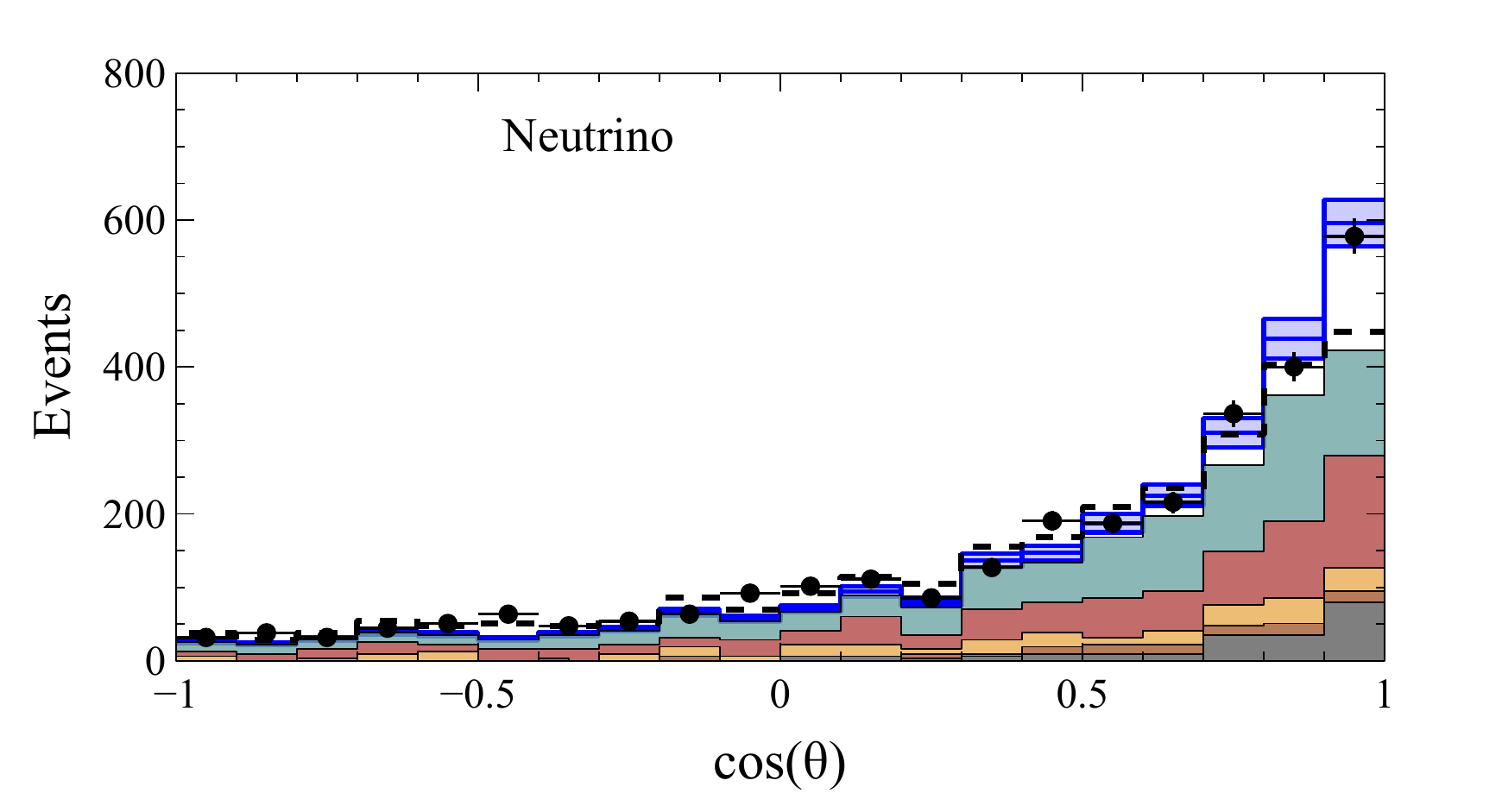}\\
\includegraphics[width=0.49\textwidth]{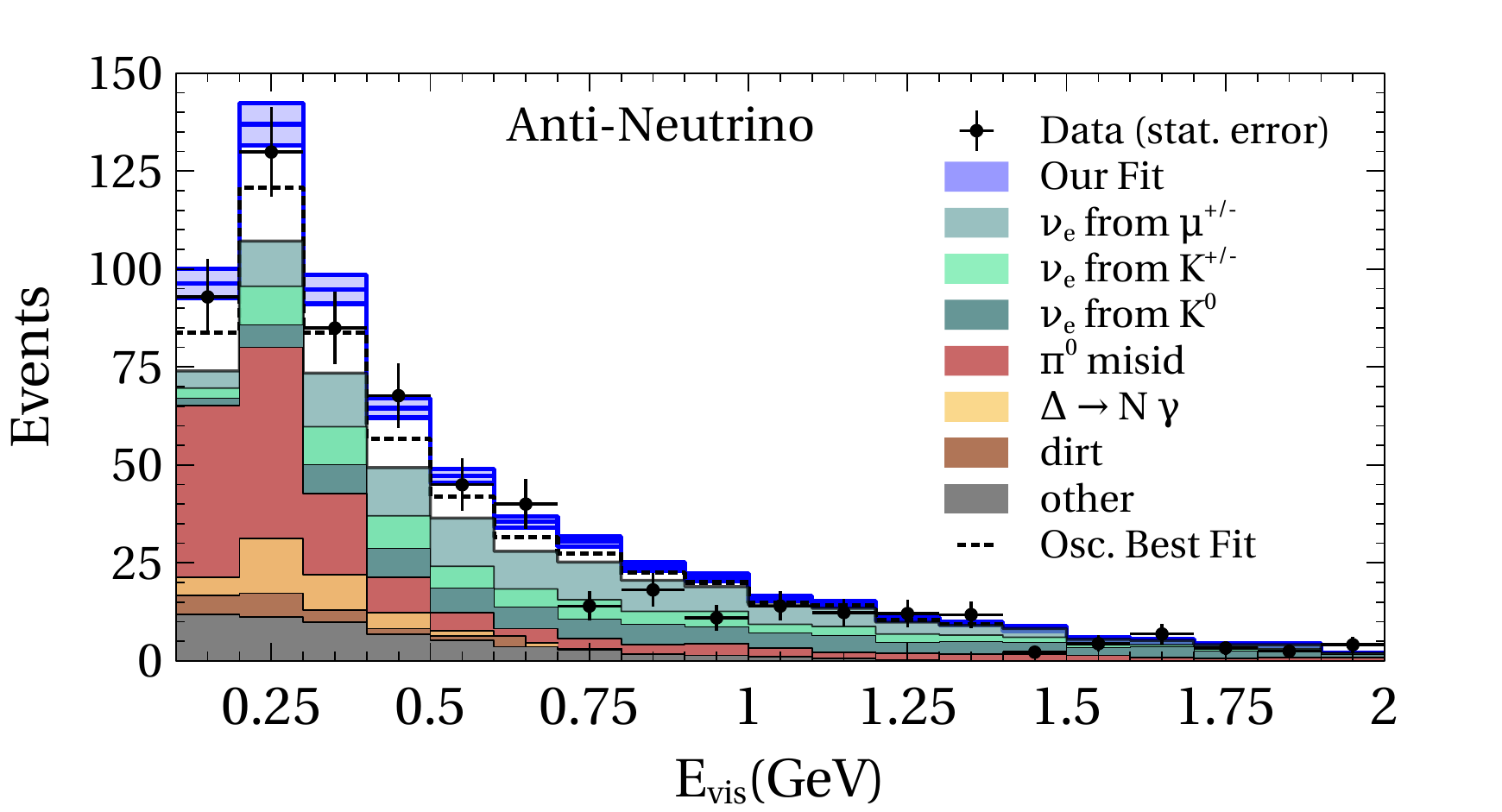}
\includegraphics[width=0.49\textwidth]{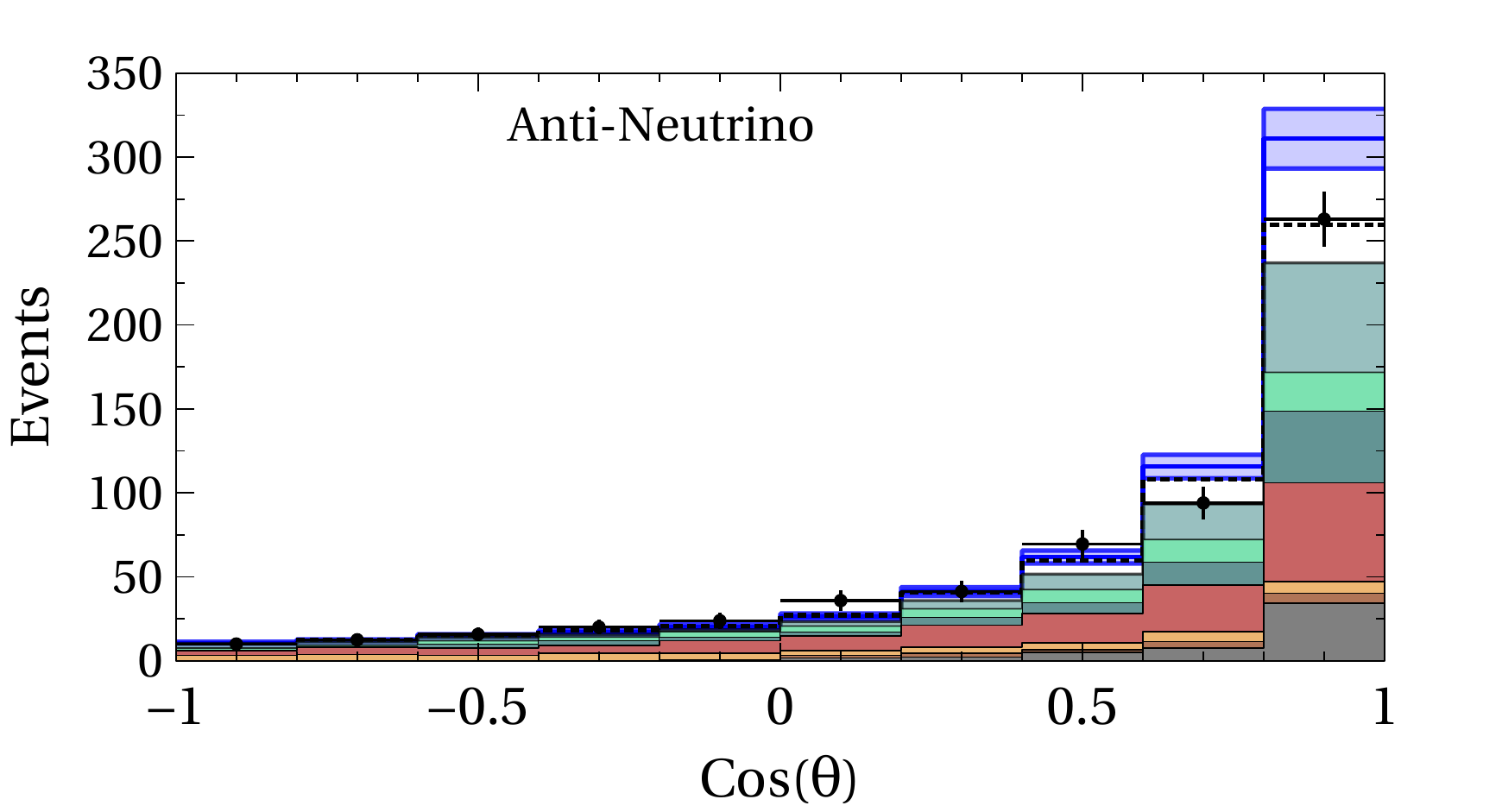}
\caption{The MB electron-like events (backgrounds and signal) from~\cite{Aguilar-Arevalo:2018gpe,Aguilar-Arevalo:2020nvw}, versus the visible energy  E$_{\rm vis}$  and versus the cosine of the emitted angle of the light, for neutrino (top)  and anti-neutrino (bottom) runs. Data points show statistical errors, whereas the blue band shows (estimated) systematic errors. The blue solid line is the prediction of our model. The parameter values used in calculating it are shown in Table~\ref{tab2}.}
\label{MB-events}
\end{figure}

Fig.~\ref{MB-events} shows, in each of the 4 panels, the data points\footnote{\vspace{-0.15cm}Note that the latest data for the neutrino mode, corresponding to $18.75 \times 10^{20}$ POT, as detailed in~\cite{Aguilar-Arevalo:2020nvw} have been used in our fit.}, SM backgrounds and  the prediction of our model (blue solid line) in each bin. Also shown (black dashed line) is the oscillation best fit. The left  panel plots show the distribution of the measured visible energy, E$_{\rm vis}$, plotted against the events for neutrinos (top) and anti-neutrinos (bottom). For our model, E$_{\rm vis}$ corresponds to $E_{h'}$. The right panels show the corresponding angular distributions for the emitted light. The benchmark parameter values used to obtain the fit from our model are shown in Table~\ref{tab2}. The plots have been prepared using fluxes, efficiencies POT exposures and other relevant information from~\cite{Aguilar-Arevalo:2018gpe,Aguilar-Arevalo:2020nvw} and references therein. We see that very good fits to the data are obtained both for energy and angular distributions. (The data points show only statistical uncertainties.). We have assumed a $15\%$ systematic uncertainty for our calculations. These errors are represented by the blue bands in the figures.
\begin{table}[htb]
\begin{center}
{\selectfont\fontsize{10.4}{10.4}{
 \begin{tabular}{|@{\hspace{0.04cm}}c@{\hspace{0.04cm}}|@{\hspace{0.04cm}}c@{\hspace{0.04cm}}|@{\hspace{0.04cm}}c@{\hspace{0.04cm}}|@{\hspace{0.04cm}}c@{\hspace{0.04cm}}|@{\hspace{0.04cm}}c@{\hspace{0.04cm}}|@{\hspace{0.04cm}}c@{\hspace{0.04cm}}|@{\hspace{0.04cm}}c@{\hspace{0.04cm}}|@{\hspace{0.04cm}}c@{\hspace{0.04cm}}|@{\hspace{0.04cm}}c@{\hspace{0.04cm}}|@{\hspace{0.04cm}}c@{\hspace{0.04cm}}|@{\hspace{0.04cm}}c@{\hspace{0.04cm}}|}
  \hline
  \makecell{$m_{\nu_4}$\\[-0.2cm] (MeV)}& \makecell{$m_{Z'}$\\[-0.2cm] (MeV)} & \makecell{$m_{h'}$\\[-0.2cm] (MeV)}& \makecell{$m_{H}$\\[-0.2cm] (MeV)}& $|U_{\mu 4}|^2$& $g_{B}$ & $g_{d}$ &$\sin\delta$&$y_{e(\mu)}^{h'}\!=\!y^{e(\mu)} \sin\delta$&$y_{e(\mu)}^{H}\!=\!y^{e(\mu)} \cos\delta$ &$|y^{e\tau} y^{\tau e}|$\\
  \hline
   50 & $800$ & $23$& $106$& $2.6\!\times\! 10^{-5}$& $3\!\times\! 10^{-4}$ & $2.85$ &0.28&$0.45(1.8)\!\times\! 10^{-4}$&$1.5(6.0)\!\times\! 10^{-4}$&$5.6\!\times\! 10^{-7}$ \\
  \hline
 \end{tabular}
\caption{Benchmark parameter values used for event generation in MB and for calculating the muon and electron $g-2$.}
\label{tab2}
}}
\end{center}
\end{table}

As mentioned earlier, the LSND observations measure the visible energy from the Cerenkov and scintillation  light of  an assumed electron-like event,  as well as the $2.2$~MeV photon resulting from coincident neutron capture on hydrogen. In our model, this corresponds to the  scattering diagrams in Fig.~\ref{FD-SP-MB} where the target is a neutron in the  Carbon nucleus. Unlike the case of  MB above, where both coherent and incoherent processes contribute to the total cross section, the LSND cross section we have used  includes only an incoherent contribution. Using the  same benchmark parameters as were used to generate the MB results, as well as all pertinent information on fluxes, efficiencies, POT etc  from~\cite{PhysRevLett.75.2650,PhysRevLett.77.3082,PhysRevC.54.2685,PhysRevC.58.2489,Aguilar:2001ty}, we find a very small excess ($1-2$ events,  from the DIF flux only), compared to the much larger observed excess reported by  LSND~\cite{Aguilar:2001ty}. We note that our calculations do not include effects arising from final state interactions or other considerations like nuclear screening or multiple scattering inside the nucleus, which could play a role at the LSND energies~\cite{1805.07378}. The KARMEN experiment similarly employed a mineral oil detection medium,  but was less than a third of the size of  LSND. It did not have a significant DIF flux, but had similar incoming proton energy and efficiencies. Unlike LSND, it saw no evidence of an excess. A simple scaling estimate using our LSND result gives  $\sim0$ events in  KARMEN using our model, which is consistent with their null result.

\begin{figure}[t!]
\center
\includegraphics[width=0.4\textwidth]{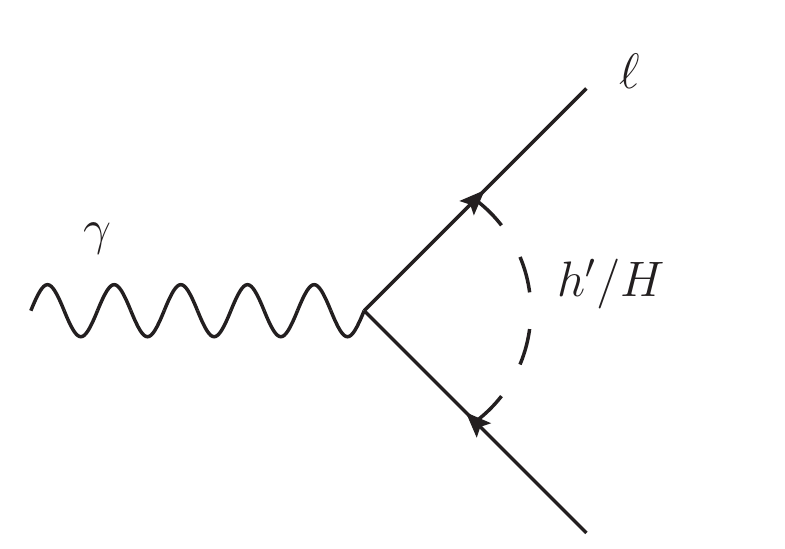}
\caption{One-loop contribution to lepton anomalous magnetic moments from the neutral scalars $h'$ and $H$.}
\label{dipoleFD}
\end{figure}
\subsection{Muon anomalous magnetic moment}
\label{sec4-B}
The one-loop contribution of a scalar $\phi$ (as shown in Fig.~\ref{dipoleFD}) to the muon anomalous magnetic dipole moment is given by~\cite{Jackiw:1972jz,Leveille:1977rc}
\begin{equation}
\Delta a^\phi_{\mu}=\frac{(y^\phi_{\mu})^2}{8\pi^2}\int_0^1 dx \frac{(1-x)^2(1+x)}{(1-x)^2+x\, r_\phi^2},
\end{equation}
where $r_\phi=m_\phi/m_\mu$, and   $\phi=h',H$. $y^\phi_\mu$ is the coupling strength of the scalar $\phi$ with the muon pair, which is defined in Eq.~(\ref{phi-ll-coupling}).

\begin{figure}[t!]
\vspace{-0.6cm}
\center
\includegraphics[width=0.6\textwidth]{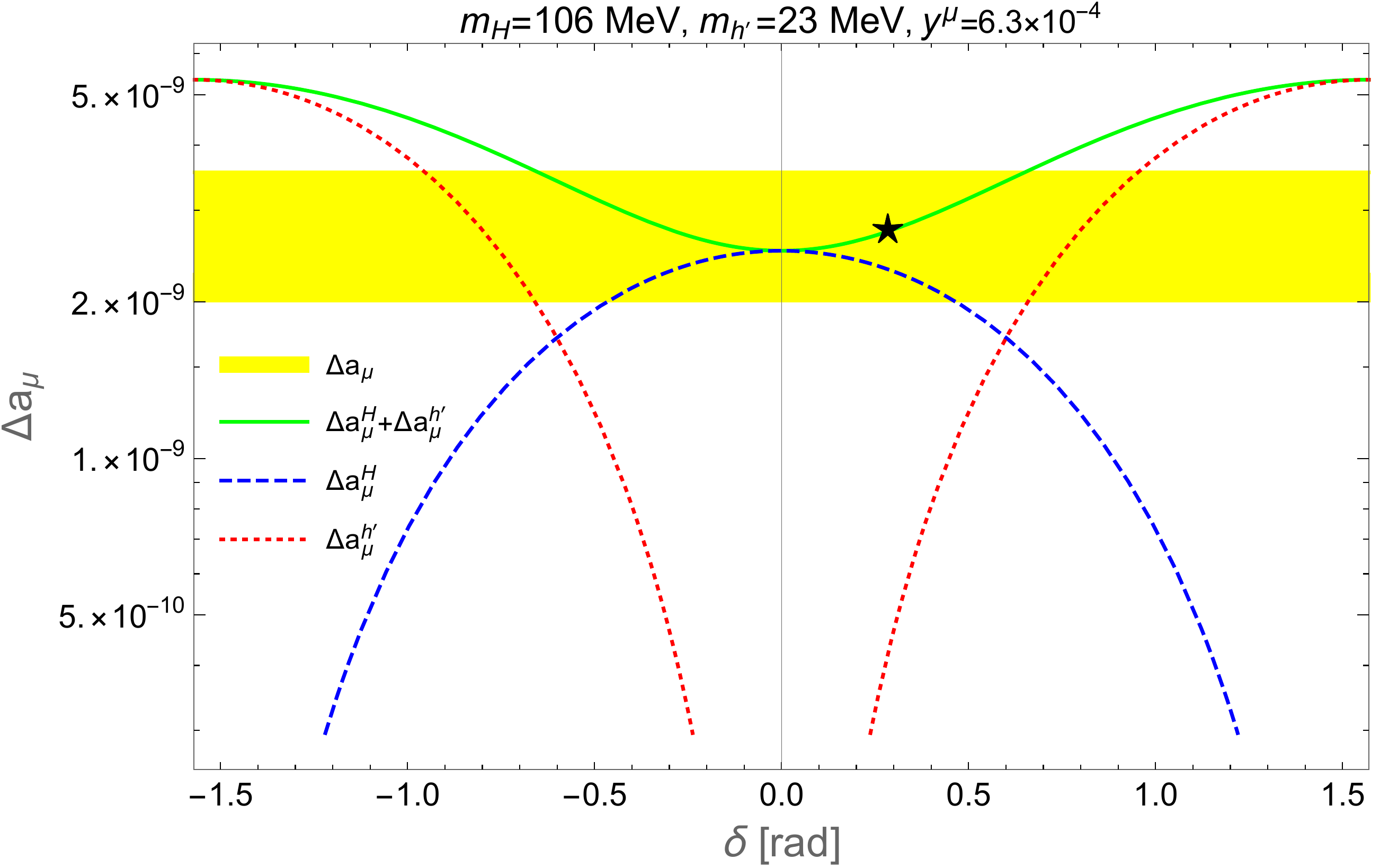}
\caption{Muon anomalous magnetic moment versus the scalar mixing angle $\delta$, along with our benchmark in Table~\ref{tab2} denoted by the black star.}
\label{deltamu-delta}
\end{figure}

In our scenario, both $h'$ and  $H$ have comparable contributions to the muon anomalous magnetic moment given that they have light masses $\leq 1$~GeV~\cite{Liu:2020qgx,Liu:2018xkx}. In Fig.~\ref{deltamu-delta}, we show the relative contributions of $h'$ and $H$ to $\Delta{a_\mu}$ as a function of the scalar mixing angle~$\delta$. The blue dashed and red dotted lines correspond to the muon anomalous magnetic moment contributions of $H$ and $h'$  ($\Delta a_\mu^{H}$ and $\Delta a_\mu^{h'}$), respectively, while the green solid line refers to their sum ($\Delta a_\mu^{H}+\Delta a_\mu^{h'}$). In addition, the horizontal yellow band indicates the $3.7\sigma$
muon $g-2$ discrepancy: $\Delta a_\mu=(2.79\pm 0.76)\times 10^{-9}$~\cite{Aoyama:2020ynm} and the black star denotes our benchmark in Table~\ref{tab2}. We note that in this figure $m_{h'}$, $m_H$ are fixed to fit the MB measurements, as discussed in the previous section. We see that both $h'$ and $H$ have reasonable and comparable contributions to the total muon anomalous magnetic moment $\Delta a_\mu$ and their ratio $\Delta a_\mu^{h'}/\Delta a_\mu^{H}\sim \tan^2\delta$.  Although in our scenario are fixed $m_{h'}$ and $m_{H}$ to fit the MB measurements, in a more general situation $y^\mu$ and the angle $\delta$ are still free parameters and one can fix them to fit the central value for $\Delta a_\mu$.

For a suitably selected combination of $y^\mu$ and $\delta$ ($y^\mu=6.3\times 10^{-4}$ and $\sin\delta=0.28$), our benchmark (denoted by the black star) is situated in the experimental allowed region (yellow band), close to  the central value for $\Delta a_\mu$ ($=2.74\times 10^{-9}$). For our benchmark, it is clearly seen that while  the total muon anomulous magnetic moment $\Delta a_\mu$ is dominated by the $H$ contribution $\Delta a_\mu^H$ (blue dashed line),  the $h'$ contribution (red dotted line) is $18\%$ of $\Delta a_\mu$,  which is not negligible. The constraints on $y^\phi_{\mu}~(\phi=h',H) $ are shown in Fig.~\ref{constraints-fig}. We see that both $y^{h'}_{\mu}$ and $y^{H}_{\mu}$  sit in the experimentally allowed region of the current  constraint of BaBar~\cite{Lees:2014xha} and the future sensitivity of Belle-II~\cite{Batell:2017kty}.
\subsection{Electron anomalous magnetic moment}
\label{sec4-C}
In this sub-section, we consider the one-loop contribution of a light scalar $\phi$ ($h',H$ in our model) to the electron anomalous magnetic moment which is given by~\cite{Jackiw:1972jz,Leveille:1977rc}
\begin{equation}
\Delta a^\phi_e=\sum_{\ell=e,\mu,\tau}\frac{y^\phi_{e\ell}\,y^\phi_{\ell e}}{8\pi^2}\int_0^1 dx \frac{(1-x)^2(r_\ell+x)}{(1-x)^2+x\, r_\phi^2+(1-x)(r_\ell^2-1)},
\end{equation}
where $r_X=m_X/m_e$ and $y^\phi_{ee}=y^\phi_e$ is the coupling strength of the scalar $\phi$ with the electron pair, as defined in Eq.~(\ref{phi-ll-coupling}). To evade the BR($\mu\to e\gamma$) and BR($\tau\to \mu\gamma$) experimental upper bounds~\cite{TheMEG:2016wtm,Aubert:2009ag}  and explain the electron $g-2$ anomaly, hereafter, we have chosen $y_{\mu e(e\mu)}^\phi$ and  $y_{\mu \tau(\tau\mu)}^\phi$ to be sufficiently tiny and the product $y^\phi_{\tau e}y^\phi_{e\tau}$ is negative. Overall, $\Delta{a_e}$ gets a positive contribution due to the non-vanishing Yukawa couplings $y^\phi_e$ which are fixed to fit the MB measurements, as discussed in section \ref{sec4-A}. Also, it gets a negative contribution from $\tau$ inside the loop, since  the product $y^\phi_{\tau e}y^\phi_{e\tau}$ is negative and is essentially a free parameter in our scenario. Thus, one can choose the absolute value of this product to fit the central value of $\Delta a_e$. Note that  $y^{h'}_{\tau e(e\tau)}=y^{\tau e(e\tau)} \sin\delta$ and $y^{H}_{\tau e(e\tau)}=y^{\tau e(e\tau)} \cos\delta$. 

\begin{figure}[t!]
\vspace{-0.6cm}
\center
\includegraphics[width=0.8\textwidth]{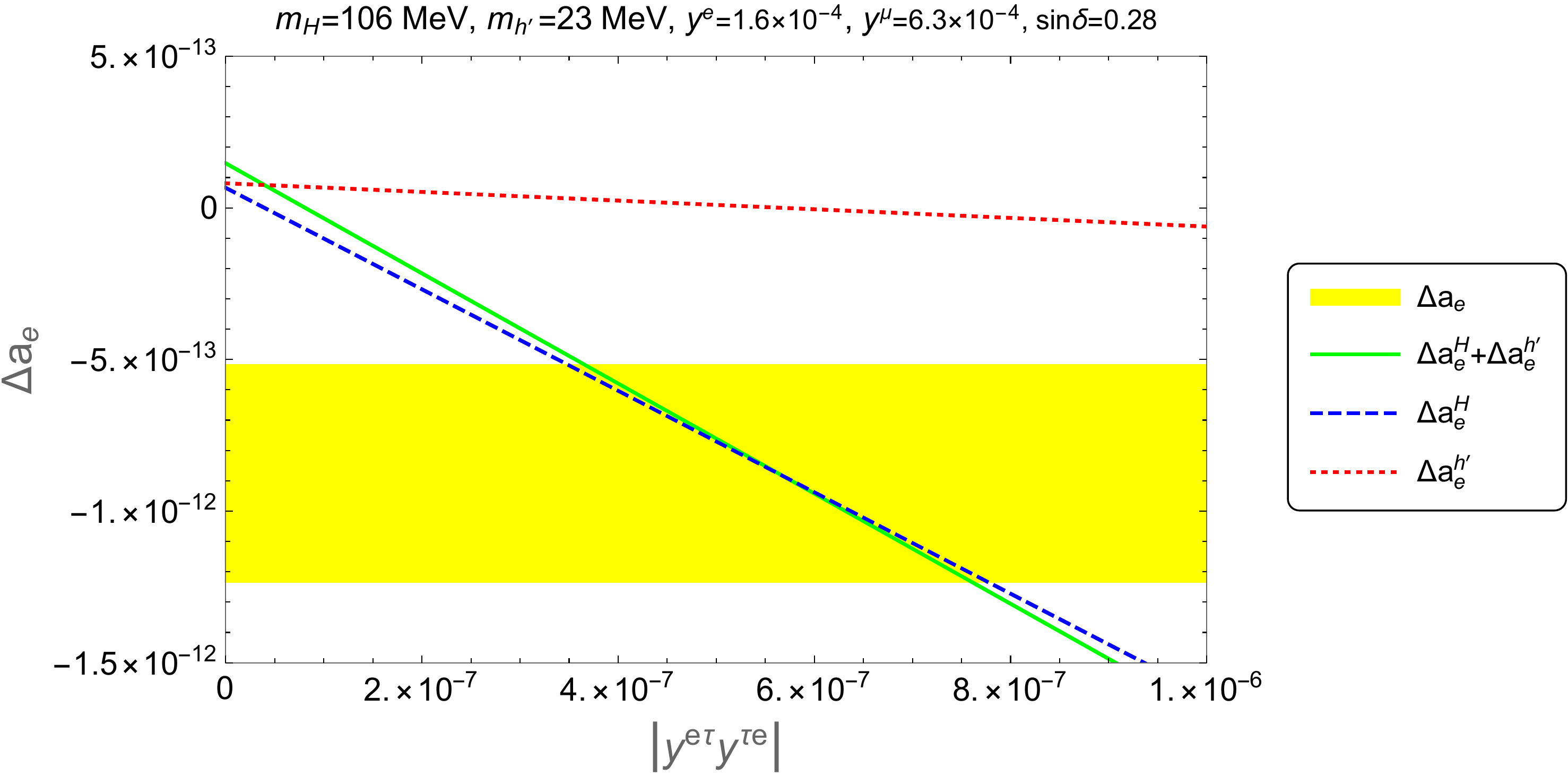}
\caption{Electron anomalous magnetic moment as a function of the absolute value of $y^{e\tau} y^{\tau e}$.}
\label{yteyet-dae}
\end{figure}

In our scenario, as mentioned earlier, $h'$ and  $H$ have light masses and consequently both contribute to the electron $g-2$ anomaly, $\Delta a_e$. In Fig.~\ref{yteyet-dae}, we show the relative contributions of $h'$ and $H$ to $\Delta{a_e}$ versus the absolute product $|y^{e\tau} y^{\tau e}|$. The blue dashed and red dotted lines correspond to the electron anomalous magnetic moment contributions of $H$ and $h'$  ($\Delta a_e^{H}$ and $\Delta a_e^{h'}$), respectively, while the green solid line refers to their sum ($\Delta a_e^{H}+\Delta a_e^{h'}$). In addition, the horizontal yellow band indicates the $2.4\sigma$ discrepancy between the experimental measurement and theoretical prediction: $\Delta a_e=(-8.7\pm 3.6)\times 10^{-13}$~\cite{Parker:2018vye}.  We see that both $h'$ and $H$ have approximately the same positive contribution ($\simeq 10^{-13}$) to the total electron anomalous magnetic moment $\Delta a_e$  at $|y^{e\tau} y^{\tau e}|=0$, which is coming from $e$ inside the loop (electron contribution).  Additionally, since  the contribution of $\tau$ inside the loop (tau contribution) which owes its sign to  the product $y^\phi_{\tau e}y^\phi_{e\tau}$ is negative,  $\Delta a_e$ gets a negative contribution overall. This originates mainly  from the $H$ contribution ($\Delta a_e^H$), as it is clearly seen from Fig.~\ref{yteyet-dae}. In this figure,  the other relevant parameters have the benchmark values shown in Table~\ref{tab2}. It is clear that for $|y^{e\tau} y^{\tau e}|\simeq 5.6\times 10^{-7}$, our benchmark sits near the central value for $\Delta a_e$~($=-8.7\times 10^{-13}$).

\section{Discussion on Constraints}
\label{sec5}
This section is devoted to  a discussion of constraints  that the proposed scenario must satisfy,  and related issues as well as future tests of the various elements of our proposal. Subsection {\bf{A}} focuses on bounds related to the additional $U(1)$ and its gauge boson and couplings, while Subsection {\bf{B}} discusses constraints related to the scalar sector extension. We have, for the most part, restricted our discussion to the regions of parameter space relevant to our scenario.

\begin{figure}[t!]
\center
\includegraphics[width=0.50\textwidth]{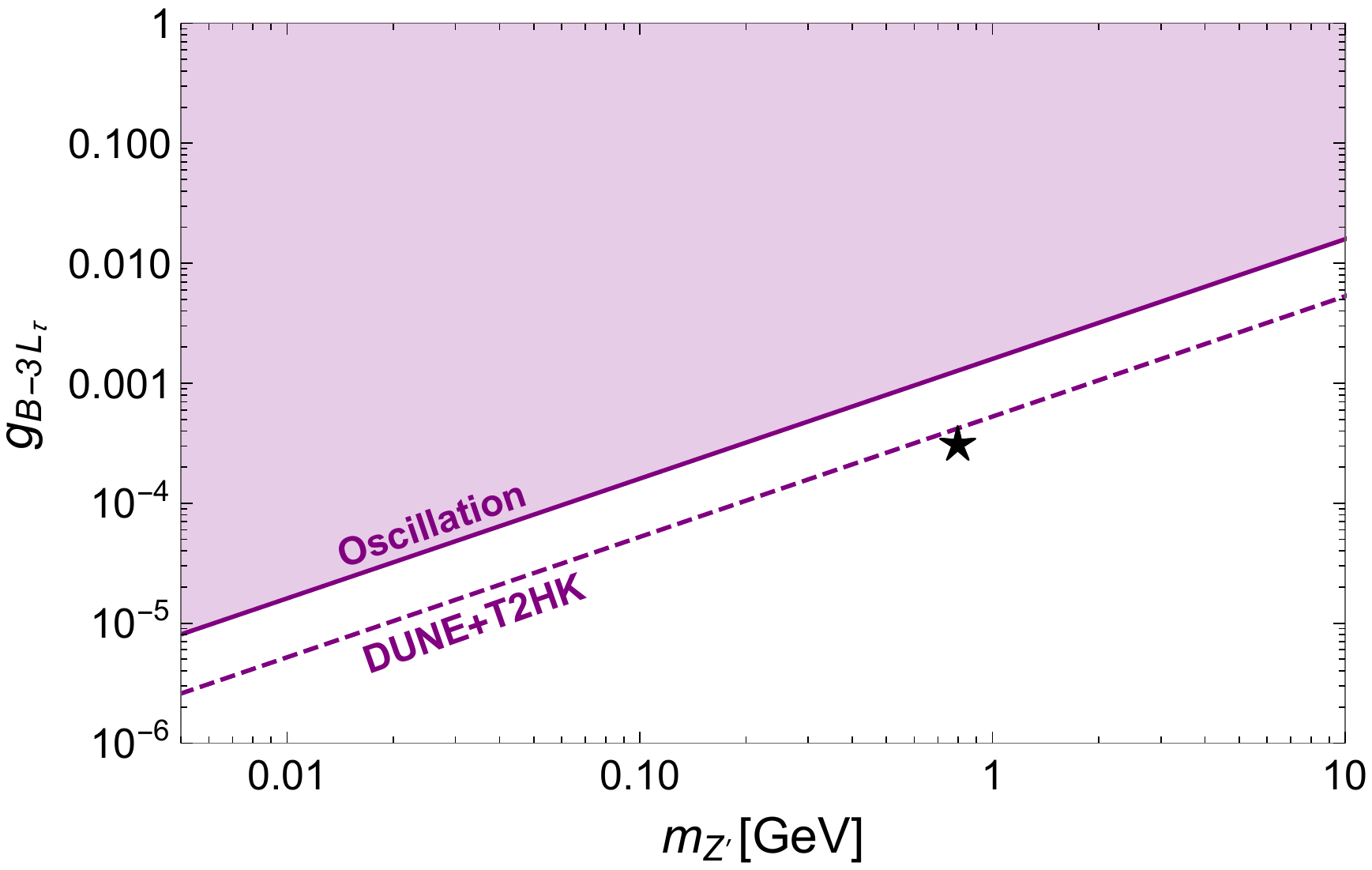}~\includegraphics[width=0.49\textwidth]{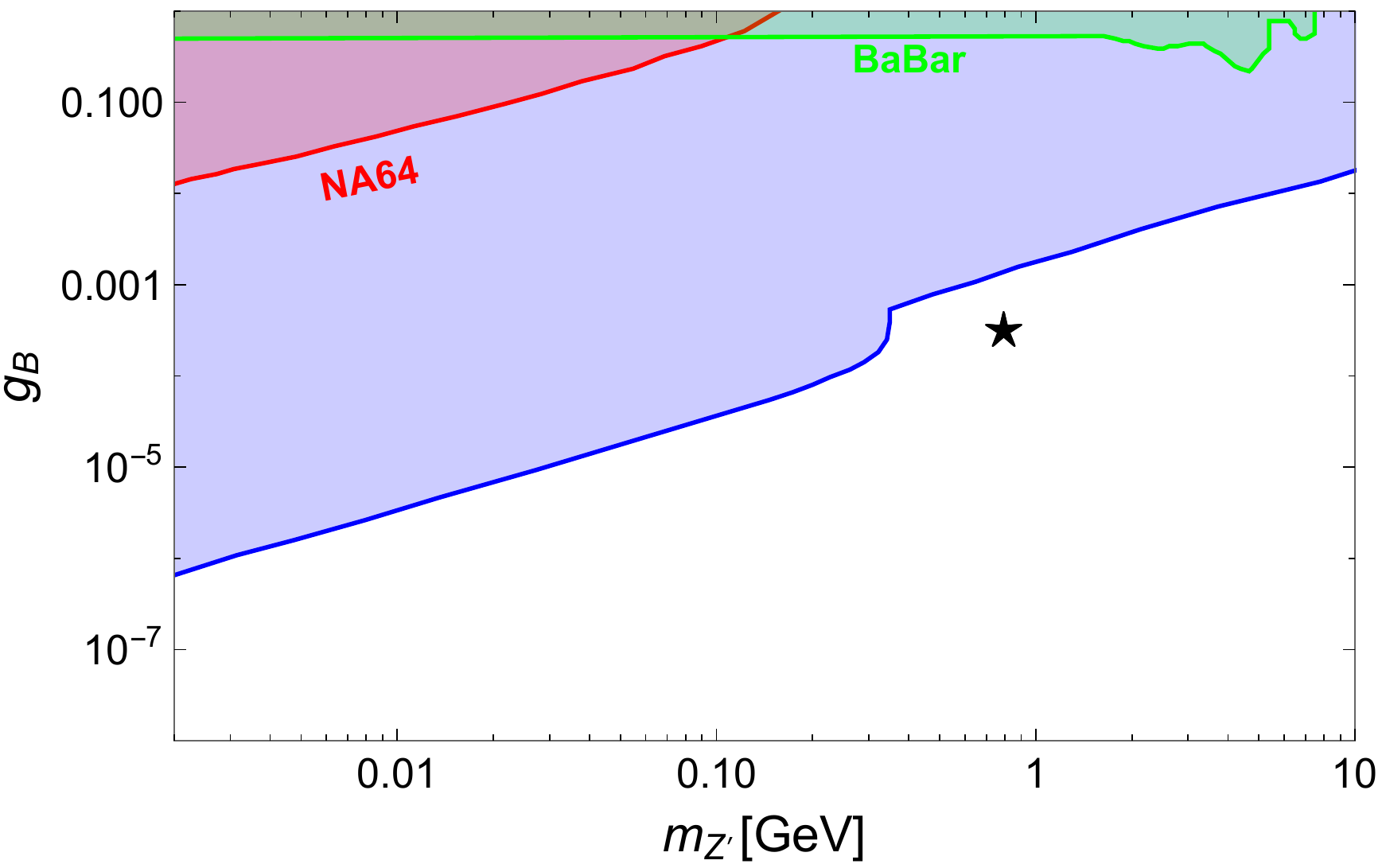}
\caption{Constraints on $m_{Z'}$ and $g_{B-3L_\tau}$ from oscillation experiments, denoted by the solid line and shaded region above it, along with projected sensitivities of T2K and DUNE (left panel, adapted from~\cite{Han:2019zkz}) and on $m_{Z'}$ and $g_B$ from NA64~\cite{Banerjee:2017hhz} and BaBar~\cite{Lees:2017lec}, along with theoretical bounds from (see text)~\cite{1705.06726, 1707.01503} (right panel, adapted from~\cite{1801.04847}) along with our benchmark in Table~\ref{tab2} denoted by the black star.}
\label{MZp-constraints-fig}
\end{figure} 
\subsection{The $U(1)$ extension}
\label{sec5-A}
\textit{\bf{Constraints on $m_{Z'}$ \textit{and}  $g_{B-3L_\tau}$:}} Strong constraints on this coupling and the associated $Z'$ mass arise from oscillation experiments as well as various decay searches~\cite{1404.4370, Farzan:2016wym, Heeck:2018nzc,Han:2019zkz}. Fig.~\ref{MZp-constraints-fig} (left panel) shows these bounds, along with our benchmark point.  We note that there is a significant difference between the bounds on the coupling coming from~\cite{Han:2019zkz} and~\cite{Heeck:2018nzc}. The reason lies in the choice, respectively,  of the LMA and LMA $+$ LMA-D and KamLAND solutions made by them. For more details the reader is referred to~\cite{Esteban:2018ppq}.
Our benchmark point is compatible with both bounds, comfortably  with~\cite{Han:2019zkz}, but only marginally so with~\cite{Heeck:2018nzc}. Future tests of these parameter values would be possible via oscillation measurements  at DUNE~\cite{Acciarri:2015uup} and T2HK~\cite{Abe:2018uyc}, as discussed in~\cite{Han:2019zkz}. Other experiments sensitive to $\tau$ interactions, like DONuT~\cite{Kodama:2007aa} and the future emulsion detectors SHiP~\cite{Anelli:2015pba}, FASER$\nu$~\cite{Abreu:2020ddv,Abreu:2019yak} and SND@LHC~\cite{Ahdida:2020evc} could provide additional constraints on the parameter space for $m_{Z'}$ and $g_{B-3L_\tau}$~\cite{Kling:2020iar}.

\textit{\bf{Constraints on $m_{Z'}$ \textit{and}  $g_B$:}}
The gauging of baryon number via a light boson associated with a $U(1)_{B}$ symmetry, which primarily interacts with quarks is subject to a number of constraints on its mass $m_{Z'}$ and the gauge coupling $g_B$~\cite{1801.04847}. Assuming that the primary modes of decay are invisible, the strongest of these come from theoretically computed bounds arising from anomaly cancellation by heavy fermions, which lead to enhanced interaction rates for processes involving the longitudinal mode of the $Z'$~\cite{1705.06726, 1707.01503}. In addition, constraints from searches by NA64~\cite{Banerjee:2017hhz} and BaBar~\cite{Lees:2017lec} for a light vector decaying to invisible become relevant.  We show these in Fig.~\ref{MZp-constraints-fig} (right panel), along with our benchmark values. 

\begin{figure}[t!]
\includegraphics[width=0.5\textwidth]{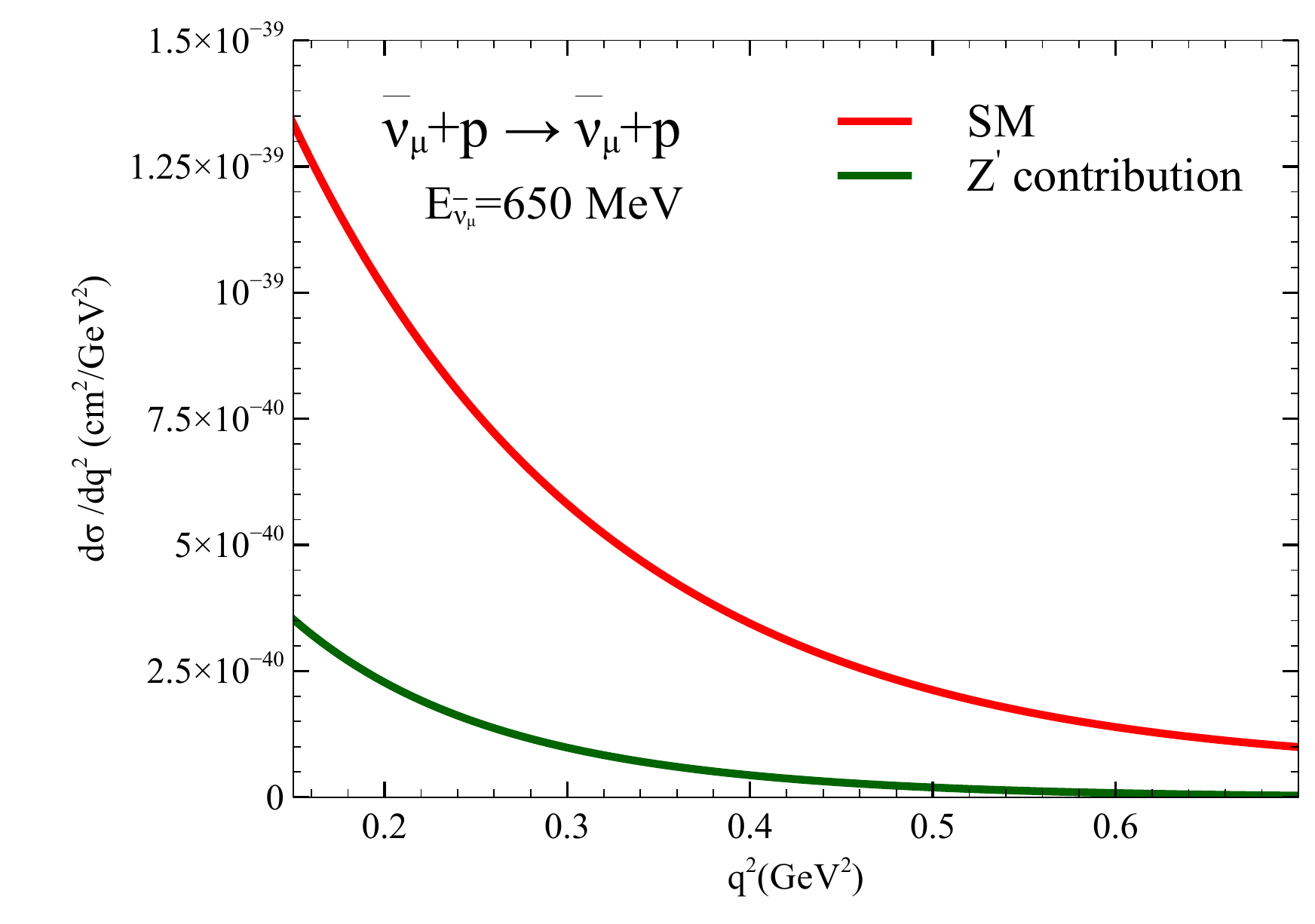}
\caption{The SM NC quasi-elastic anti-neutrino-proton cross section, compared to the contribution  obtained from our model with the $Z'$ due to its couplings to baryons and to $\nu_d$. The parameter values used in calculating the $Z'$ contributions  are the same as those used for our MB result, and are given in Table~\ref{tab2}.}
\label{QE}
\end{figure}
\textit{\bf{Contributions to NC $\nu$-nucleon scattering at both low and high energies:}} At low energies, an  important constraint arises from NC quasi-elastic neutrino-nucleon scattering, to which the new $Z'$ would contribute
via an amplitude proportional to $g_d\,g_B\, U_{\mu4}$.  MB has measured this cross section in the relevant range~\cite{0909.4617}. Fig.~\ref{QE} shows the SM differential cross section for muon anti-neutrino scattering  and compares it to the cross section from our model. We see that the contribution from the 
latter  stays safely below the SM anti-neutrino cross section, which, of course, is lower than that for neutrinos and thus provides a more conservative basis for comparison. We note that our process with the $Z'$ mediator does not distinguish between neutrino and anti-neutrino scattering, unlike the SM case. It also adds $10-25\%$ to the SM cross section, over the range shown. Interestingly,  MB NC measurements have been  fitted with an axial mass $M_A$ which is significantly higher than the value from the global average value of this parameter, indicating that the measured cross section is higher than expected, with one possible conclusion being that it is receiving contributions from new physics.

\begin{figure}[t!]
\includegraphics[width=0.50\textwidth]{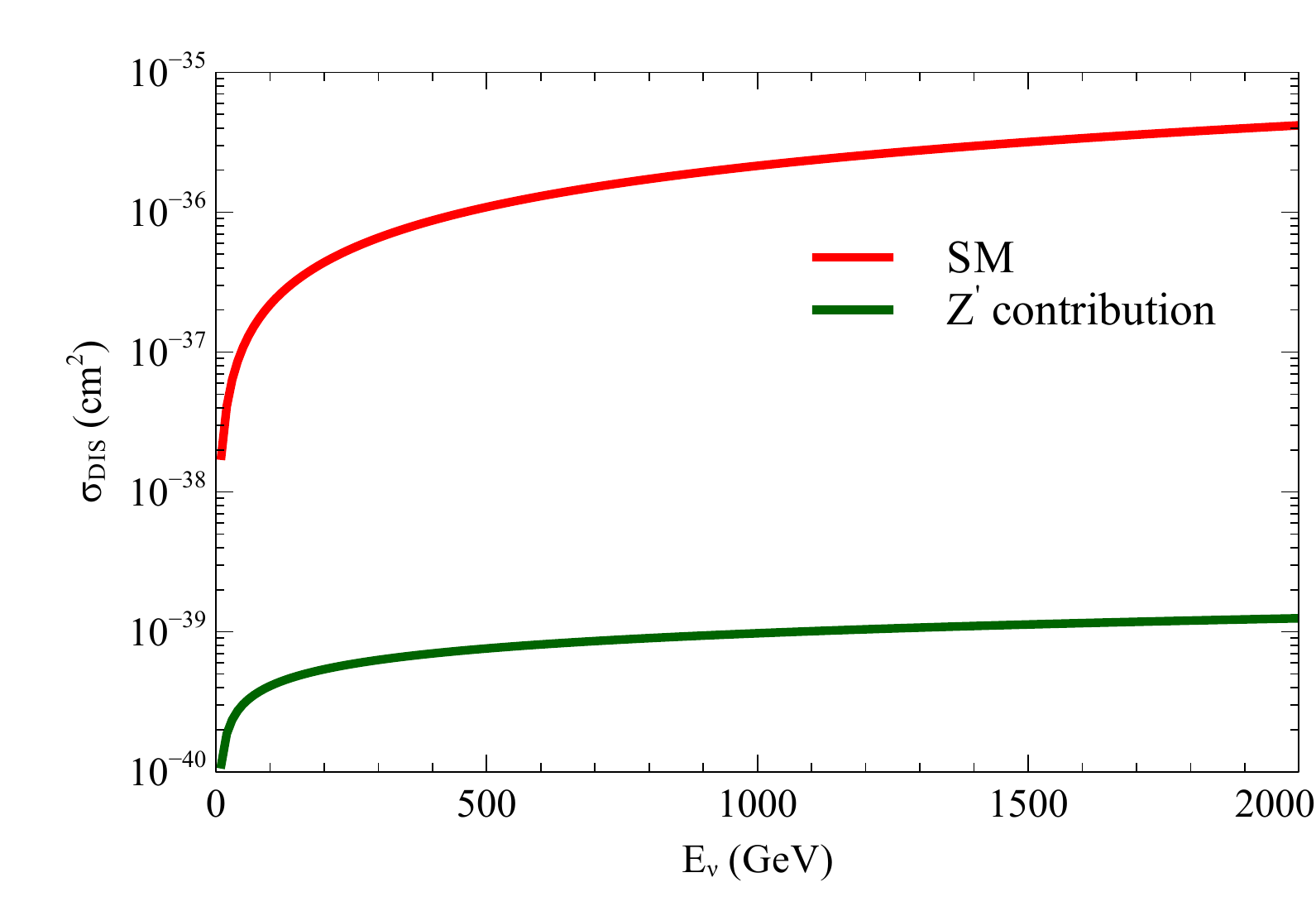}
\caption{The SM NC DIS cross section, compared to that obtained from our model. The parameter values used in calculating the $Z'$ contribution are given in Table~\ref{tab2}.}
\label{DIS}
\end{figure}
IceCube and DeepCore are a possible laboratory for new particles which are produced via neutrino nucleon scattering~\cite{Coloma:2017ppo, Coloma:2019qqj}.
Fig.~\ref{DIS} shows our check for  contributions of the model to deep inelastic scattering (DIS), comparing it to the SM total NC cross section for $\numu$-nucleon scattering. The $Z'$ contributions are more than three orders of magnitude lower. We note that the DeepCore and IceCube detectors would be sensitive to the new particles  and the interaction in our model in two ways: $a)$ by a possibly measurable increase in the neutrino nucleon NC event rate, and $b)$ via the decay  of $h'$ into an $e^+e^-$ pair if, after its production in a NC event mediated by $Z'$, it travels a distance long enough to signal a double bang event (about 10~m in DeepCore, and $\sim$ a few hundred m in IceCube). The lifetime of the $h'$ in our scenario is $c\tau\sim 10^{-4} $~m. The distances travelled even at very high energies are much smaller than the resolution necessary to signal a double bang event. In addition, as Fig.~\ref{DIS} shows, the high energy NC cross section  stays several orders of magnitude below the SM cross section. We note that similar to the low-energy case above, the $Z'$ contribution has been calculated taking into account the enhancement it receives due to $g_d\, U_{\mu4}$ at the neutrino vertex.

\textit{\bf{Constraints on $m_{\nu4}$, $|U_{\mu4}|$ and $|U_{e4}|$:}} The mass of the dark neutrino in our model has a wider possible range than that in scenarios where it is required to decay inside the MB detector~\cite{1009.5536, 1807.09877, 1808.02915} to obtain the electron-like signal.  Its main role here is that of a portal connecting the SM neutrinos via mixing to the $Z'$. Nonetheless, heavy sterile neutrino masses and mixings are tightly constrained by a number of experiments, as well as astrophysics and cosmology,  and these bounds are discussed and summarized in~\cite{Bolton:2019pcu,0901.3589,1011.3046,1110.1610,
1502.00477,1511.00683,1605.00654,1904.06787,1909.11198}. We assume that the $\nu_d$ does not constitute an appreciable fraction of DM in the universe, and has dominantly  invisible decay modes. Our benchmark value for its mass is $\sim 50$~MeV, and this along with the mixings we assume are in conformity with the existing bounds.

\textit{\bf{Constraints from NOMAD:}} The NOMAD experiment carried out  a  search for neutrino induced single photon events at high energies, $E_\nu\sim 25$~GeV~\cite{Kullenberg:2011rd}. It  obtained an upper limit of $4.0 \times 10^{-4}$ single photon events for every  $\numu$ induced  charged-current event. Clearly, extrapolating our calculations to NOMAD energies would  be invalid, given that the calculational procedures we use to obtain the pair production contributions do not apply there.  NOMAD used coherent pion kinematics with one photon to arrive at their bound. We then examine the ratio of  the cross section for our process, including coherent effects,  to the charged current total inclusive incoherent muon production cross section measured by NOMAD at $E_\nu\sim 25$~GeV, and obtain a ratio below the upper bound given by NOMAD.

\textit{\bf{Constraints from CHARM II and MINERVA:}} We find that in our model,  the $Z'$ does contribute to the neutrino electron scattering cross section at these detectors to the extent of about $\sim 14\%$, leading to a very  mild tension with their observations when flux and other uncertainties are accounted for. 

\textit{\bf{Constraints from CE${\nu}$NS:}} Any additional $U(1)$ with a vector $Z'$ mediator that couples to neutrinos and baryons could conceivably receive large contributions from coherent elastic neutrino-nucleon  scattering (CE${\nu}$NS)~\cite{PhysRevD.9.1389,Kopeliovich:1974mv},  since it would receive an enhancement proportional to the square of the number of nucleons.  In our scenario, in spite of the choice of gauge groups being $U(1)_{B-3L\tau}$ or $U(1)_B$, the $Z'$ does effectively couple to muon neutrinos (Fig.~\ref{FD-SP-MB}). The amplitude for this process receives an  added enhancement from the fact that the effective active neutrino-$Z'$ coupling is $g_d\,U_{\mu 4}$, which can be significantly larger than $g_B$. 

The COHERENT Collaboration~\cite{Akimov:2017ade} has recently observed CE${\nu}$NS, for neutrinos in the energy range of $16-53$~MeV, and concurrently set stringent bounds on the parameters $g_B$ and $m_{Z'}$. The values of $g_B$ and $m_{Z' }$ chosen by us respect these constraints, but the coupling for the amplitude of the enhanced process,  $g_B\,g_d\,U_{\mu 4}$ does not. However, the neutrino beam energies in COHERENT are  below the kinematic range required for the process in Fig.~\ref{FD-SP-MB}, since besides nuclear/nucleon recoil, a heavy neutrino of mass $\sim 50$~MeV must be produced in the final state. Thus the event rate in COHERENT remains unaffected by our scenario.

\begin{figure}[t!]
\center
\includegraphics[width=0.47\textwidth]{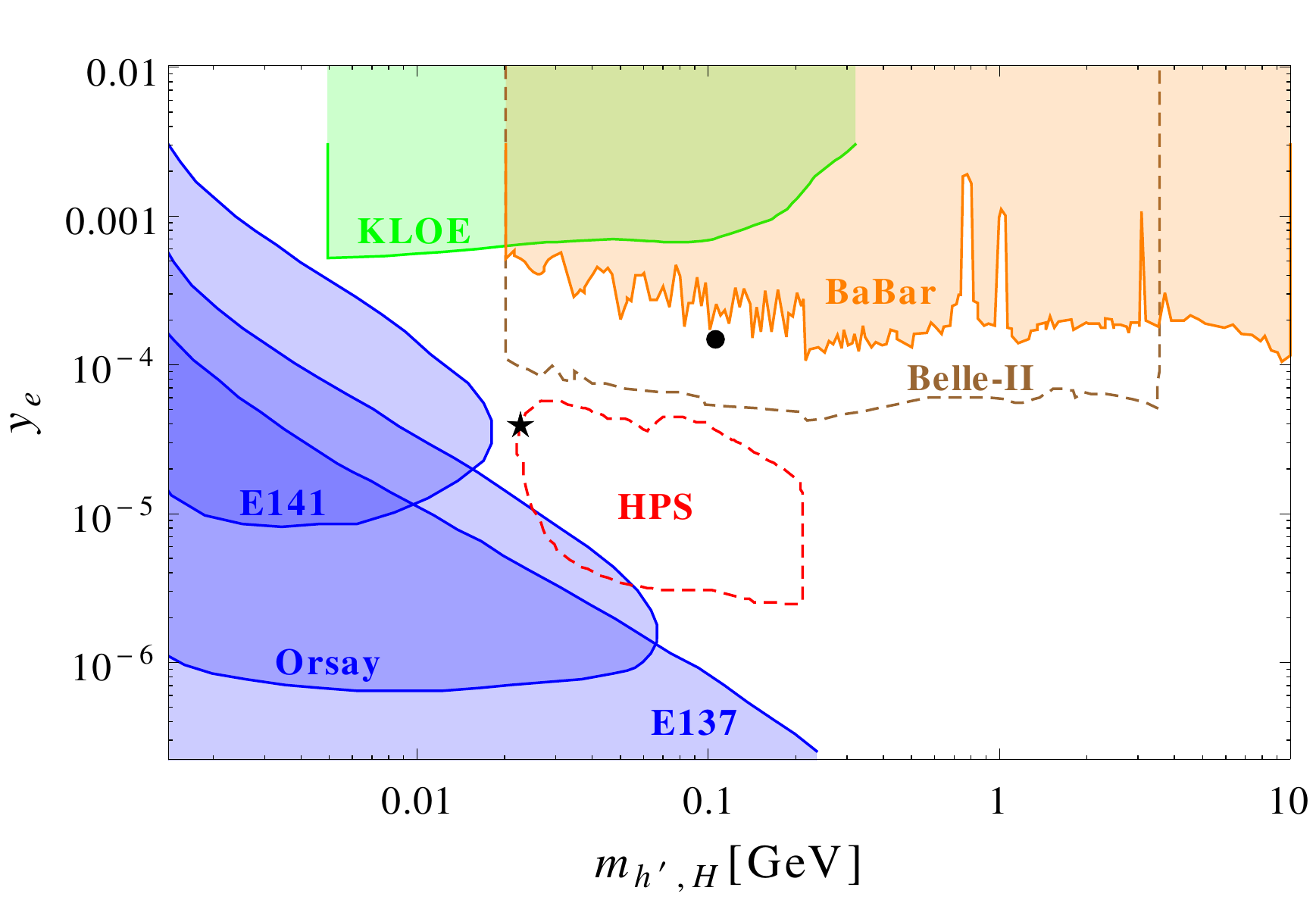}~~\includegraphics[width=0.5\textwidth]{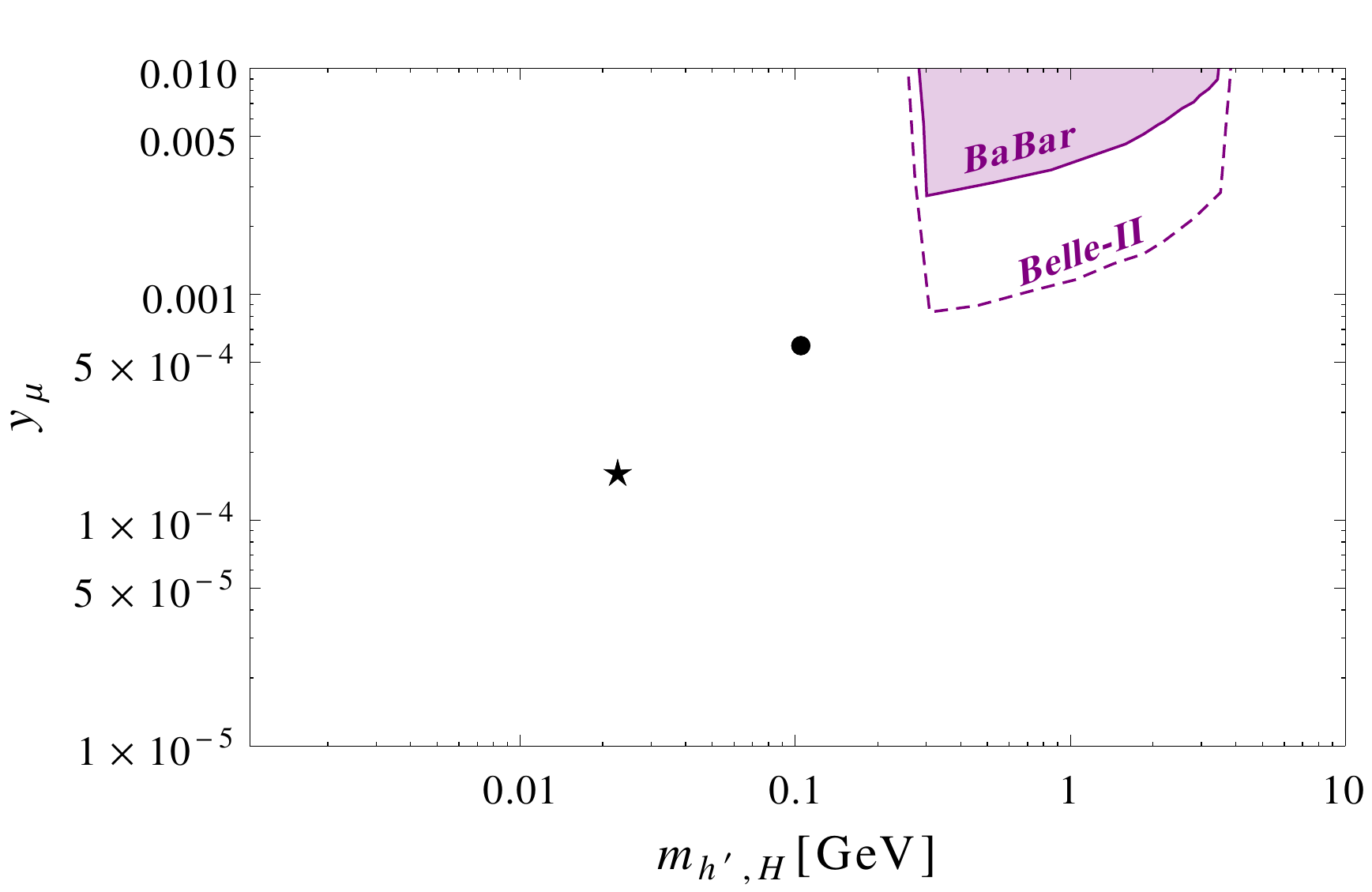}
\caption{Relevant constraints to our scenario, where the color shaded regions with solid boundary indicate to the  excluded regions by current experiments, and the unshaded regions with dashed boundaries are future projections (see text for details).}
\label{constraints-fig}
\end{figure}
\subsection{The extended scalar sector}
\label{sec5-B}
\textit{\bf{Constraints on $y_e$ and $m_{h'}$ from dark photon searches:}} A dark photon search looks for its decay to lepton pair. These bounds can be translated~\cite{Alves:2017avw, Knapen:2017xzo} to constraints on a light scalar which couples to leptons. We show these translated constraints relevant to our scenario from KLOE~\cite{Anastasi:2015qla}, BaBar~\cite{Lees:2014xha} and the projected future sensitivity from Belle-II in Fig.~\ref{constraints-fig} (left panel)~\cite{Batell:2017kty}.

\textit{\bf{Constraints on $y_e$ and $m_{h'}$ from electron beam dump experiments:}} A light scalar with couplings  to electrons could be searched for~\cite{Liu:2016qwd, Batell:2016ove} in beam dump experiments via its decay to an $e^+ e^-$ pair or photons. Relevant to the mass range under consideration here are the experiments E137~\cite{Bjorken:1988as}, E141~\cite{Riordan:1987aw} and ORSAY~\cite{Davier:1989wz}. The forbidden regions are shown in Fig.~\ref{constraints-fig} (left panel). In the future, the HPS fixed target experiment~\cite{Battaglieri:2014hga} which will scatter electrons on tungsten, will be able to constrain the displaced decays of a light scalar. Its projected sensitivity is  also shown in this figure.

\textit{\bf{Constraints from ND280:}} As discussed in~\cite{Brdar:2020tle}, the T2K near-detector, ND280, is in a position to provide bounds on new physics related to the MB LEE. Relevant to our work here, the specific decay $h'\rightarrow e^+e^-$ could be observable in the  Ar TPC associated with this detector. In our model, however, this decay is prompt, hence the Ar gas must act as both target and detection medium if this is to be observed. Since the target mass is only 16 kg, however, the number of events is unobservably small in our case.

\textit{\bf{Future tests of the muon and the electron $g-2$:}} The E989 experiment~\cite{Grange:2015fou} at Fermilab  is soon  likely to announce results of measurements of the muon $g-2$ which will have significantly higher precision than current measurements. This will be complemented by measurements  of this quantity at comparable precision by an experiment at J-PARC and the E34 Collaboration~\cite{Abe:2019thb}. An important supplementary effort will be the measurement of the hadronic contributions to the muon magnetic moment by the MUonE experiment~\cite{Abbiendi:2016xup} at CERN, which will  determine them at  uncertainties below those in present theoretical calculations. Finally, continuing and improved measurements of the fine structure constant are likely to determine the future significance of the discrepancy in the electron $g-2$.

\textit{\bf{Constraints on $y_\mu$ and $m_H$ from colliders:}} BaBar has provided constraints~\cite{Batell:2016ove, Batell:2017kty} on these parameters via their search for $e^+ e^-\rightarrow \mu^+\mu^-\phi$, where $\phi$ is a generic light scalar. Also shown in Fig.~\ref{constraints-fig} (right panel) is the future projection for Belle-II~\cite{Batell:2017kty}. Our benchmark points, as shown, are below these bounds. 

\textit{\bf{Constraints on $y_\tau$ and $m_H$ from BaBar:}} Very recently, BaBar has provided strong constraints~\cite{BABAR:2020oaf} on the parameter $\xi$, which is the ratio of the effective coupling ($y_{e,\mu,\tau}$ in our model) of a light scalar to a fermion compared to its SM Yukawa coupling $(m_f/v)$. BaBar looks for narrow width decays of a leptophilic scalar $\phi$, produced radiatively from $\tau$-lepton via  $e^+e^-\rightarrow\tau^+\tau^-\phi$, followed by $\phi\rightarrow e^+e^-$. In our case, noting that $y_{e,\mu,\tau}$ are all independent,  this translates to a bound on $y_\tau$ and $m_{H}$. For our mass range for $H$, of $100-150$~MeV, this implies that $y_\tau$ remain below $\approx 3.5\times10^{-3}$.  In our scenario, $y_\tau$, is independent and essentially free, and can be kept small. We also note that the independence of $y_\mu$ from $y_\tau$ in our scenario ensures that the bound on $y_\tau$ from BaBar does not automatically translate into a bound on $y_\mu$, unlike the case where $\xi$ is the same for all leptonic generations.

\textit{\bf{Constraints from $\tau\to e\gamma$ :}} In our calculation for  $\Delta a_{e}$, the   BR($\tau\to e\gamma$) has a non-zero value due to the non-vanishing Yukawa couplings of the $\phi$-$e$-$\tau$ interactions, $y^\phi_{\tau e(e\tau)}$. We thus calculated  this BR($\tau\to e\gamma$)~\cite{Lavoura:2003xp} mediated by the light scalars $h'$ and $H$ using  $|y^{e\tau} y^{\tau e}|\simeq 5.6\times 10^{-7}$. We find that this yields a total BR$(\tau\to e\gamma) \simeq 5.4\times 10^{-10}$. We note that this is very small compared with the experimental upper bound,  BR$(\tau\to e\gamma) <1.1\times 10^{-7}$~\cite{Aubert:2009ag}, and hence is not a concern.

\textit{\bf{Constraints from Higgs physics:}}. We note that in  the model considered here, the $h$ is almost identical to the  SM Higgs, with negligible mixings to the other neutral scalars ($h'$ and $H$). This makes the constraints from Higgs observations not a matter of immediate~concern.

\textit{\bf{Stability of the scalar potential:}} We have examined the behaviour of the potential as the fields tend to infinity, in order to ensure it is stable. Our benchmark parameters satisfy the vacuum stability conditions. The details are provided in the Appendix~\ref{App:AppendixB}.

\textit{\bf{Collider constraints on the heavy charged CP-even scalars $H^\pm$:}} Drell-Yan processes at both LEP and the LHC can produce pairs of the $H^\pm$, which can subsequently decay to a neutrino and a lepton each. Bounds set on supersymmetric particles~\cite{Sirunyan:2018vig, Aad:2014yka, Sirunyan:2018nwe} which would mimic these final states  can be translated to bounds on $H^\pm$, as discussed in~\cite{Babu:2019mfe, Jana:2020pxx}. These lead to a lower  bound on the charged scalar mass of $m_{H^\pm}> 110$~GeV.

\textit{\bf{Electro-weak precision constraints on the heavy charged CP-even scalars $H^\pm$ and pseudoscalar $A$:}} The oblique parameters $S, T$ and $U$ are a measure of the effects new particles can have on gauge boson self energies. The effects of scalars in an expanded Higgs sector  on these parameters have been discussed in~\cite{Funk:2011ad, Grimus:2007if, Grimus:2008nb}. For models in the alignment limit, bounds using the $T$ parameter are particularly significant in constraining the plane of mass differences between $a)$ the SM-like Higgs and the charged $H^\pm$, and $b)$ the  $H^\pm$ and the pseudoscalar~\cite{Jana:2020pxx, Babu:2018uik}. Essentially, one finds that either the masses of the pair in $a)$ or that in $b)$ need to be close to each other, while the other mass difference can be large, $\eg\sim$ a few hundred GeV. In our scenario, if the dominant contribution to $\Delta a_\mu$ is to originate from an $H$ with a mass below $200$~MeV, one is led to the mass hierarchy $m_A\sim m_{H^\pm}$ $\gg m_H$.

\section{Summary and concluding remarks}
\label{sec6}
Among several anomalous signals at low energy experiments, the MB LEE and the discrepancy in the measured value of the anomalous magnetic moment of the muon stand out,  due to their statistical significance, the duration over which they have been present and the scrutiny and interest they have generated. Our effort in this paper  takes the viewpoint that these anomalies are  due to interlinked underlying new physics involving a new $U(1)$ connecting the SM and the dark sector.

Pursuant to this, starting with the MB LEE, we find that a light $Z' $ vector portal associated with $U(1)_{B-3L_\tau}$, which is anomaly-free, or a  $U(1)_B$ extension of the SM, combined with a second Higgs doublet allows a very good fit to the excess. The $Z'$ obtains its mass from a dark sector singlet scalar, and is coupled to a dark neutrino.  The Higgs sector thus comprises of three CP-even scalars, $h$, which is predominantly  SM Higgs-like, and  $h'$ and $H$ which are  light compared to $h$ and the charged Higgses of the model.  $h' , H$ are coupled both  to the dark sector and  to SM fermions via mixing. In MB, the $h' (H) $ is produced via the $Z'$-$h' (H)$-$Z'$ coupling and decays primarily to an $e^+e^-$ pair.  Both $h'$ and $H$ contribute to both the MB LEE and the muon and electron $g-2$, but for our choice of benchmarks, the $h' (H)$ contributes dominantly to the MB LEE (the muon and electron $g-2$).

Our work underscores the role light scalars may play in understanding  low energy anomalies that persist and survive further tests, and the possibility that a light $Z'$ may provide an important portal to the dark sector. This $Z'$ need not be unique as long as it couples in a flavor universal way to quarks. The couplings to leptons are constrained to be very small, however, especially  for the first two generations. Overall, we provide a template for a model with an additional $U(1)$ that agrees very well with MB data while staying in conformity with all known constraints.

We note that  a singlet scalar mass-mixed with the SM Higgs along with the $Z'$,  could, in principle  have provided an economical solution for the MB LEE. However, the fermionic couplings of such a scalar are constrained to be very tiny and cannot be used to generate the MB excess. This motivates the need  for a second Higgs doublet mixed with the dark sector. We find that when incorporated, the interplay of the scalars via mixing allows us to understand both the MB signal and the observed anomalous value of  the muon  magnetic moment in a manner that satisfies existing constraints.  

\section*{Acknowledgements} 
We are thankful to Richard Hill for discussions and collaboration in the early stages of this work. We thank William Louis and Tyler Thornton for help with the making of Fig.~\ref{MB-events}. RG would like to extend special thanks to Boris Kayser, William Louis and  Geralyn Zeller for many very helpful  discussions on MB and LSND results, and to Steven Dytman, Gerald Garvey, Sudip Jana and Lukas Koch for  providing especially helpful clarifications over email. He is also grateful to Gauhar Abbas, Ismail Ahmed, N. Ananthanarayan, K. S. Babu,  Andr\'e de Go\"uvea, Jeff Dror, Rikard Enberg, Chris Hearty, Robert Lasenby, John LoSecco, Pedro Machado, Tanumoy Mondal, Biswarup Mukhopadhyaya, Satya Mukhopadhyay,  Roberto Petti, Santosh Rai, S Uma Sankar and Ashoke Sen for helpful discussions and email communications. He thanks Patrick deNiverville, Suprabh Prakash and Sandeep Sehrawat for  assistance in the early stages of this work. He is grateful to  the Theory Division and the 
 Neutrino Physics Center at Fermilab for  hospitality and visits where this work benefitted from discussions and a conducive environment. SR thanks KM Patel for useful discussions. He is grateful to Fermilab, where this work was initiated,  for support via the Rajendran Raja Fellowship. WA, RG and SR also acknowledge support from the XII Plan Neutrino Project of the Department of Atomic Energy and the High Performance Cluster Facility at HRI (http://www.hri.res.in/cluster/). 
\appendix
\section*{Appendices}
\section{Diagonalization of CP-even Higgs mass matrix}
\label{App:AppendixA}
In the basis $\left(H_1^0, H_2^0, H_3^0\right)$,
the mass matrix of the neutral CP-even Higgses is given by
\begin{equation}
m^2_{\cal H} = \left(
\begin{array}{ccc}
\lambda_1 v^2 &\lambda_6 v^2  &  \lambda'_3 v v'\\
\lambda_6 v^2 & \bar{m}_H^2 & \lambda'_5 v v' \\
 \lambda'_3 v v' & \lambda'_5 v v' &2 \lambda'_2 v'^2\end{array}
\right).
 \end{equation} 
Now if $\lambda_6 \simeq 0  \simeq \lambda'_3$, then we get the alignment limit
\textit{i.e.} one of the CP-even Higgs mass eigenstates aligns with the vev
direction of the scalar field. In the alignment limit, the mass matrix becomes
\begin{equation}
m^2_{\cal H} \simeq \left(
\begin{array}{ccc}
\lambda_1 v^2 & 0  &  0\\
0 & \bar{m}_H^2 & \lambda'_5 v v' \\
 0 & \lambda'_5 v v' &2 \lambda'_2 v'^2\end{array}
\right).
 \end{equation}
Now,
\begin{eqnarray}
{Z^{\cal H}} m^2_{\cal H}  (Z^{\cal H})^T&=& ({\cal M}^2)^{\rm diag}={\rm
diag}\{m_{h}^2,m_{H}^2,m_{h'}^2\},
\end{eqnarray}
where
\begin{equation}
{Z^{\cal H}}=\left(
\begin{array}{ccc}
 1 & 0 & 0 \\
 0 & \cos\delta & -\sin\delta \\
 0 & \sin\delta & \cos\delta \\
\end{array}
\right),~~{\rm with}~~\tan{2 \delta}=\frac{-2 \lambda'_5 v
v'}{\bar{m}_H^2-2 \lambda'_2 v'^2}.
\end{equation}
The eigenvalues of the mass matrix are
\begin{equation}
m^{2}_{h,H,h'}\simeq\left\{\lambda_1 v^2,\frac{1}{2}\left(\bar{m}_H^2+2 \lambda'_2 v'^2\pm \sqrt{(\bar{m}_H^2-2
\lambda'_2 v'^2)^2 + 4 (\lambda'_5 v v')^2 }\right)\right\}.
\end{equation}
If we choose $\bar{m}_H^2 = (102~\rm{MeV})^2$, $\lambda'_5 v v' = -(53.8~\rm{MeV})^2$ and $2 \lambda'_2 v'^2 = (37~\rm{MeV})^2$, we get $m_H = 106$~MeV, $m_{h'}=23$~MeV and $\sin \delta =0.28$, which fit our benchmark in Table~\ref{tab2}.
\section{Vacuum Stability}
\label{App:AppendixB}
For a stable vacuum, the potential should be bounded from below as the field strength approaches to infinity from any directions. In this limit, only the quartic part of the potential is relevant. In the alignment limit ($\lambda_{6}\simeq 0 \simeq \lambda'_3 $) and for simplicity we consider  $\lambda_{7} = 0= \lambda'_4 $. With those considerations, the quartic part of the potential becomes
\begin{eqnarray}
V_4 &=& \frac{\lambda_1}{2} (\phi_{h}^\dagger\phi_{h})( \phi_{h}^\dagger\phi_{h}) + \dfrac{\lambda_2}{2} (\phi_{H}^\dagger\phi_{H})( \phi_{H}^\dagger\phi_{H})+\lambda_3 (\phi_{h}^\dagger\phi_{h})( \phi_{H}^\dagger\phi_{H})+\lambda_4 (\phi_{h}^\dagger\phi_{H})(\phi_{H}^\dagger\phi_{h})\nonumber\\
&+&\frac{\lambda_5}{2}( (\phi_{h}^\dagger\phi_{H})^{2} + (\phi_{H}^\dagger\phi_{h})^{2}) + \lambda'_{2} (\phi_{h'}^\ast\phi_{h'})^2 +\lambda'_5 (\phi_{h'}^\ast\phi_{h'})(\phi_{h}^\dagger\phi_{H} + \phi_{H}^\dagger\phi_{h}).
\label{V4-VS}
\end{eqnarray}
We can parameterize the fields as~\cite{ElKaffas:2006gdt}
\begin{equation}
|\phi_{h}|=r c_\vartheta s_\varphi,~~~~~|\phi_{H}|=r s_\vartheta s_\varphi,~~~~~
|\phi_{h'}|=r c_\varphi,~~~~~\phi_{h}^\dagger\phi_{H} = |\phi_{h}||\phi_{H}| \,  \rho \, e^{i \gamma},
\end{equation}
where $s_x\equiv \sin x$, $c_x\equiv \cos x$, $r\geq 0, \, \, \vartheta \in [0,\pi/2], \,\, \varphi \in [0, \pi/2], \,\, \rho \in [0,1]$ and $\gamma \in [0, 2\pi]$. The potential can be written as
\begin{eqnarray}\nonumber
\dfrac{V_4}{r^4} &=&\left[\dfrac{\lambda_1}{2} c^4_\vartheta  + \dfrac{\lambda_2}{2} s^4_\vartheta + \lambda_3 s^2_\vartheta c^2_\vartheta + \lambda_4 s^2_\vartheta c^2_\vartheta \rho^2 + \lambda_5 s^2_\vartheta c^2_\vartheta \rho^2 c_{2 \gamma}\right] s^4_\varphi  \\
&+& \lambda'_2 c^4_\varphi + 2 \, \lambda'_5 \rho c_\vartheta s_\vartheta c^2 \phi s^2_\varphi c_\gamma.
\label{V4-P}
\end{eqnarray}
In our case, $\lambda_4$ is negative and other terms containing $\rho$ are function of phase $\gamma$. Hence, we consider $\rho=1$. Now,
\begin{eqnarray}
( \lambda_5 s^2_\vartheta c^2_\vartheta  c_{2 \gamma} s^4_\varphi + 2 \, \lambda'_5 c_\vartheta s_\vartheta c^2_\varphi s^2_\varphi c_\gamma  )_{\rm min} &>& (-|\lambda_5| s^2_\vartheta c^2_\vartheta s^4_\varphi
- 2 |\lambda'_5| c_\vartheta s_\vartheta c^2_\varphi s^2_\varphi).
\end{eqnarray}
We define
\begin{eqnarray}
\dfrac{\tilde{V}_4}{r^4} &= &\left[ \dfrac{\lambda_1}{2} c^4_\vartheta  + \dfrac{\lambda_2}{2} s^4_\vartheta + \lambda_3 s^2_\vartheta c^2_\vartheta + \lambda_4 s^2_\vartheta c^2_\vartheta - |\lambda_5| s^2_\vartheta c^2_\vartheta\right] s^4_\varphi+ \lambda'_2 c^4_\varphi - 2 \, |\lambda'_5| c_\vartheta s_\vartheta c^2_\varphi s^2_\varphi.
\label{V4-tilde}
\end{eqnarray}
Now, $\tilde{V}_4 >0$ implies that $V_4>0$. We first calculate the values of $\tilde{V}_4/r^4$ at the boundary points in the $(\vartheta, \, \, \varphi)$ plane:
\begin{equation*}
\dfrac{\tilde{V}_4}{r^4} \left(\vartheta=0, \, \varphi=\dfrac{\pi}{2}\right)= \dfrac{\lambda_1}{2}>0,~~~
\dfrac{\tilde{V}_4}{r^4} \left(\vartheta=\dfrac{\pi}{2}, \,\varphi=\dfrac{\pi}{2}\right)= \dfrac{\lambda_2}{2}>0,~~~
\dfrac{\tilde{V}_4}{r^4} (\varphi=0)= \lambda'_2 >0,
\end{equation*}
\begin{equation}\nonumber
\dfrac{\tilde{V}_4}{r^4} \left(\varphi=\dfrac{\pi}{2}\right)=  \dfrac{\lambda_1}{2} c^4_\vartheta  + \dfrac{\lambda_2}{2} s^4_\vartheta + (\lambda_3 + \lambda_4 - |\lambda_5|) s^2_\vartheta c^2_\vartheta >0.
\end{equation}
Therefore, the vacuum stability conditions can be written as
\begin{eqnarray}
\lambda_1, \lambda_2, \lambda'_2 >0,
\end{eqnarray}
and
\begin{eqnarray}
\lambda_3 + \lambda_4 - |\lambda_5| > - \sqrt{\lambda_1 \lambda_2}.
\end{eqnarray}
Also, we have to show that $\tilde{V}_4/r^4 >0$ in the interior points $(\vartheta, \varphi)$, \textit{i.e.}
\begin{eqnarray}
\label{ine-max}
 - 2 \, |\lambda'_5| c_\vartheta s_\vartheta \,  > \, -\left[ \dfrac{\lambda_1}{2} c^4_\vartheta  + \dfrac{\lambda_2}{2} s^4_\vartheta +  (\lambda_3 + \lambda_4 - |\lambda_5|) s^2_\vartheta c^2_\vartheta\right]\tan^2 \varphi
- \dfrac{\lambda'_2}{\tan^2\varphi}.
\end{eqnarray}
Maximizing the right hand side of the inequality (\ref{ine-max}) with respect to $\varphi$, we get
\begin{eqnarray}
-|\lambda'_5| c_\vartheta s_\vartheta  \, > \, - \sqrt{\lambda'_2 \left(\dfrac{\lambda_1}{2} c^4_\vartheta  + \dfrac{\lambda_2}{2} s^4_\vartheta +  (\lambda_3 + \lambda_4 - |\lambda_5|) s^2_\vartheta c^2_\vartheta\right)}.
\end{eqnarray}
Thus, we get the final condition for a stable vacuum as
\begin{eqnarray}
(\lambda_3 + \lambda_4 - |\lambda_5|) \lambda'_2 - |\lambda'_5|^2 > -\lambda'_2 \sqrt{\lambda_1 \lambda_2}.
\end{eqnarray}
\vspace{-1cm}
\bibliographystyle{apsrev}
\bibliography{NU-bib}
\end{document}